\documentclass[%
reprint,
superscriptaddress,
 amsmath,amssymb,
 aps,
 pra,
]{revtex4-2}

\usepackage{graphicx}
\usepackage{dcolumn}
\usepackage{bm}
\usepackage{braket}
\usepackage[colorlinks=true, linkcolor=blue, citecolor=blue, urlcolor=blue]{hyperref}
\usepackage{xcolor}
\usepackage{soul}
\usepackage{float}
\usepackage{cleveref}
\begin{document}

\preprint{APS/123-QED}

\title{The Role of Entanglement in Quantum Reservoir Computing with Coupled Kerr Nonlinear Oscillators}

\author{Ali Karimi}
\email{ali.karimi2@ucalgary.ca} 
\affiliation{Department of Physics and Astronomy, University of Calgary, Calgary, Alberta, Canada}
\affiliation{Institute for Quantum Science and Technology, University of Calgary, Calgary, Alberta, Canada}
\affiliation{Hotchkiss Brain Institute, University of Calgary, Calgary, Alberta, Canada}

\author{Hadi Zadeh-Haghighi}
\affiliation{Department of Physics and Astronomy, University of Calgary, Calgary, Alberta, Canada}
\affiliation{Institute for Quantum Science and Technology, University of Calgary, Calgary, Alberta, Canada}
\affiliation{Hotchkiss Brain Institute, University of Calgary, Calgary, Alberta, Canada}
\affiliation{QNeura, Calgary, Alberta, Canada}

\author{Youssef Kora}
\affiliation{Department of Physics and Astronomy, University of Calgary, Calgary, Alberta, Canada}
\affiliation{Institute for Quantum Science and Technology, University of Calgary, Calgary, Alberta, Canada}
\affiliation{Hotchkiss Brain Institute, University of Calgary, Calgary, Alberta, Canada}
\affiliation{Department of Physics, University of New Brunswick, Fredericton, New Brunswick, Canada}

\author{Christoph Simon}
\email{christoph.simon@ucalgary.ca}
\affiliation{Department of Physics and Astronomy, University of Calgary, Calgary, Alberta, Canada}
\affiliation{Institute for Quantum Science and Technology, University of Calgary, Calgary, Alberta, Canada}
\affiliation{Hotchkiss Brain Institute, University of Calgary, Calgary, Alberta, Canada}

\date{\today}

\begin{abstract}
Quantum Reservoir Computing (QRC) uses quantum dynamics to efficiently process temporal data. In this work, we investigate a QRC framework based on two coupled Kerr nonlinear oscillators, a system well-suited for time-series prediction tasks due to its complex nonlinear interactions and potentially high-dimensional state space. We explore how its performance in forecasting both linear and nonlinear time-series depends on key physical parameters: input drive strength, Kerr nonlinearity, and oscillator coupling, and analyze the role of entanglement in improving the reservoir’s computational performance, focusing on its effect on predicting non-trivial time series. Using logarithmic negativity to quantify entanglement and normalized root mean square error (NRMSE) to evaluate predictive accuracy, individual parameter sweeps show that optimal performance occurs at moderate but non-zero entanglement. Furthermore, an aggregated binned analysis reveals that this moderate entanglement is consistently associated with the optimal average predictive performance across the parameter space, an observation that persists up to a threshold in the input frequency. This relationship persists under some levels of dissipation and dephasing. In particular, we find that higher dissipation rates can enhance performance. These findings contribute to the broader understanding of quantum reservoirs for high performance, efficient quantum machine learning and time-series forecasting.

\end{abstract}
\maketitle

\maketitle

\section{\label{sec:intro}Introduction}

Modern machine learning has achieved remarkable success in processing complex datasets, yet handling temporal or sequential data—such as speech, sensor signals, or chaotic time series—remains challenging. Recurrent neural networks were designed to capture temporal dependencies, but their training is often difficult and resource-intensive due to issues like vanishing and exploding gradients \cite{bengio} and large parameter counts \cite{beyond}. This has motivated alternative approaches to temporal processing that reduce training costs. One such framework is reservoir computing (RC), which employs a fixed nonlinear dynamical system—the “reservoir”—to project inputs into a high-dimensional space, requiring training only a simple linear readout \cite{TANAKA2019100,PhysRevResearch.3.013077}. RC has proven effective for tasks like time-series prediction and chaos modeling, and has been physically realized in hardware ranging from electronic circuits to optical and photonic systems \cite{Paquot2012ua,Duport2012ve,brunner,vandoorne,Du,Kudi,PhysRevX.7.011015}.

In parallel, the past decades have seen rapid advances in quantum information science, with the promise of quantum computers solving certain problems exponentially faster than classical ones \cite{Nielsen_Chuang_2010,preskill}. However, current quantum devices in the noisy intermediate-scale quantum (NISQ) era face severe noise and decoherence limitations \cite{PhysRevResearch.3.013077}. An intriguing strategy is to combine ideas from noise-tolerant classical computing with quantum hardware \cite{PhysRevResearch.3.013077}. Quantum Reservoir Computing (QRC) was proposed as a way to exploit the natural dynamics of quantum systems for machine learning \cite{PhysRevApplied.8.024030}, analogous to classical RC but potentially leveraging quantum effects. 

Fujii and Nakajima first introduced the concept of QRC in 2017 \cite{PhysRevApplied.8.024030}, demonstrating that an ensemble of just 5–7 qubits could emulate the performance of a classical recurrent neural network with hundreds of neurons on chaotic time-series tasks. This result suggested that the large state space of even small quantum systems could be harnessed for computation. Subsequent works have explored QRC implementations on various platforms: for example, networks of qubits (spin-1/2 systems) have been used to process temporal data with high memory capacity \cite{Kutvonen2020}, and even real quantum processors with inherent noise have been treated as reservoirs, where the unavoidable decoherence can serve as a useful computational resource rather than an obstacle \cite{Suzuki2022}. QRC has also been extended to quantum tasks; for instance, a quantum reservoir processor was shown to recognize entanglement in quantum input states and estimate properties like entropy or logarithmic negativity, tasks that ordinarily would require multiple quantum measurements \cite{Ghosh2019-uf}. It has been argued that QRC can naturally achieve a form of nonlinear computation even when the underlying quantum evolution is linear, either through measurement-induced nonlinearities \cite{PhysRevResearch.3.013077} or through clever input encodings \cite{govia2021nonlinear}. 

Quantum reservoir computing with bosonic modes has attracted increasing attention. Prior studies have shown that Kerr nonlinearities can endow a quantum reservoir with rich dynamics. The intrinsic nonlinearity in this kind of systems ensures complex, nonlinear transformations of inputs \cite{Milburn,Kitagawa,Haroche2006-js}, which is crucial for reservoir computing. Moreover, Kerr oscillator systems are experimentally feasible \cite{Kanao}. Govia et al. \cite{PhysRevResearch.3.013077} proposed QRC using a single Kerr oscillator and demonstrated that the enlarged bosonic Hilbert space can improve time-series prediction performance. Kalfus et al. \cite{Kalfus} similarly showed that increasing the Hilbert-space dimension of a single-mode reservoir yields clear performance gains over classical counterparts. Networks of coupled oscillators have also been explored: Dudas et al. \cite{Dudas2023-bj} implemented QRC using two coherently coupled superconducting resonators, demonstrating that a pair of Fock-populated oscillators can process temporal data. Cheamsawat and Chotibut \cite{Cheamsawat2025-xx} analyzed how different drive and coupling regimes in a pair of coupled Kerr cavities affect information encoding. 

A key question driving QRC research is what intrinsic quantum features can provide computational advantages over classical reservoirs. Quantum systems offer two notable features: (i) a state space that grows exponentially with the number of constituents, and (ii) uniquely quantum phenomena such as superposition and entanglement. The former means that a moderately sized quantum reservoir can explore a vast space of configurations, which in principle could embed very complex mappings of the input data \cite{G_tting_2023}. The latter – entanglement and related non-classical correlations –may signal the presence of richer dynamics and memory than any classical counterpart \cite{G_tting_2023}. 

While some studies show that quantum reservoir computing can function without strongly non-classical states \cite{Nokkala2021-ed}, others suggest that genuine quantum resources—such as entanglement, coherence, or quantum chaos—may offer enhanced computational power. For example, memory capacity in qubit-based reservoirs has been found to correlate with entanglement, and recent work showed that this advantage becomes more prominent at higher input frequencies \cite{G_tting_2023,yousef}. 
However, it remains unclear if these findings hold for continuous-variable systems or if they are specific to spin-networks and state-injection encoding mechanisms. Notably, evidence suggests that more “quantumness” is not necessarily better. For instance, studies have shown that in certain tasks, such as NARMA, introducing squeezing (a form of nonclassicality) can reduce performance \cite{Garcia-Beni2024-hn}. These findings underscore that while quantum effects can be beneficial, their impact on reservoir computing remains subtle and task-dependent—especially the role of entanglement, which is still not fully understood.

To address these open questions, we investigate a QRC framework based on two coupled Kerr nonlinear oscillators. We employ Hamiltonian encoding, consistent with previous oscillator-based QRC implementations~\cite{PhysRevResearch.3.013077,Dudas2023-bj,Kalfus}, rather than the state injection techniques often utilized in spin-network memory capacity studies \cite{yousef}. This strategy is particularly advantageous for forecasting as it circumvents the experimental overhead of repeated state preparation and enables highly nonlinear feature mapping. Furthermore, it avoids the control overheads associated with resetting the reservoir’s memory \cite{minimalhamil}, ensuring that entanglement is not totally destroyed each time. Moreover, we do time-series forecasting both on linear multi-frequency signals and nonlinear NARMA time-series, which is substantially different than memory tasks because forecasting necessitates learning the signal's underlying dynamics rather than merely storing past inputs. Our approach employs both single-parameter sweeps and a binned analysis on the corresponding aggregated data. Through these methods, we uncover that moderate levels of entanglement coincide with optimal forecasting performance, up to a certain threshold in the input frequency. No such threshold was observed in the case of spin-network QRC~\cite{yousef}.

We further expand the analysis of noise tolerance by explicitly modeling dephasing alongside dissipation. Through a sweep of physical parameters—including drive strength, nonlinearity, and coupling—we analyze the interplay between entanglement and prediction error in the fully quantum regime.

Our results identify Kerr system operating regimes that optimize prediction performance and clarify that there is a nuanced relation between entanglement and optimized performance in the studied continuous-variable system.

The remainder of this paper is organized as follows: In section II, we introduce the physical model of the coupled Kerr oscillators and we define how input signals are fed into the oscillators, how outputs are read out, and how performance is evaluated. We also discuss the measures of entanglement and performance. Following that, section III presents and discusses the results of our simulations. Finally, we outline our conclusions in section IV. 

\begin{figure*}
\includegraphics[width=0.95\textwidth]{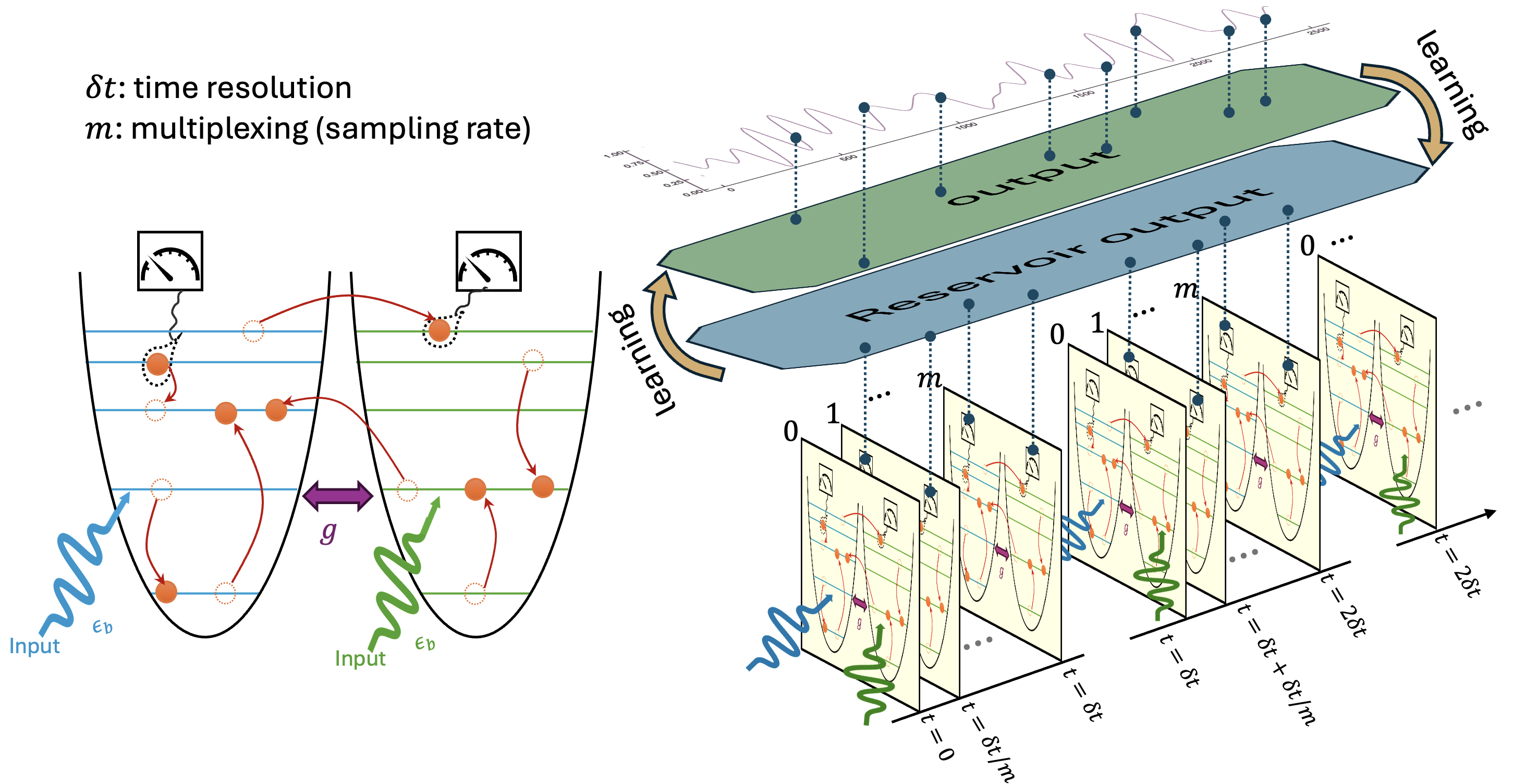}
\caption{\label{fig:scheme} \textbf{Overview of the coupled Kerr Oscillator QRC.} The oscillators are driven by the input signal with amplitudes $\epsilon_a$ and $\epsilon_b$ at a coupling strength of $g$. The reservoir state is measured in the Fock basis (read-out) at moments of $t=\delta t/m$ at a constant rate (sampling via multiplexing). In response, the quantum state of the system evolves, and the computational nodes evolve due to the natural dynamics of the system and input signal. The output of the reservoir is fed to the ridge regression to generate weights. The weights are then used to generate the output time series, which is a prediction of the target.}
\end{figure*}

\section{Model and Methodology}
Here we consider a reservoir composed of two coupled Kerr-nonlinear oscillators, $a$ and $b$ (Fig.~\ref{fig:scheme}). In this case, the reservoir "neurons" are given by the Fock basis states $\ket{n_a,n_b}$ and the reservoir output by their occupation probabilities \cite{Dudas2023-bj}. 

\subsection{Physical Model}

The system can be described by the following Hamiltonian:

\begin{equation}
\hat{H}(t)=\hat{H}_{nl} + \hat{H}_{int} + \hat{H}_{drive}
\end{equation}
where 
\begin{equation}
\hat{H}_{nl}=\hat{H}_{nl}^{(a)} + \hat{H}_{nl}^{(b)}= K_a \frac{\hat{N}_a^2}{2}+K_b \frac{\hat{N}_b^2}{2},
\end{equation}

\begin{equation}
\hat{H}_{int}=g(\hat{a}\hat{b}^\dagger + \hat{a}^\dagger \hat{b}),
\label{eq:hint}
\end{equation}

\begin{equation}
\hat{H}_{drive}=i\epsilon_a\sqrt{2\kappa_a}(\hat{a}-\hat{a}^\dagger)+i\epsilon_b\sqrt{2\kappa_b}(\hat{b}-\hat{b}^\dagger)
\label{eq:hdrive}
\end{equation}
$\hat{H}_{nl}$ describes Kerr nonlinearity where $K_a$and $K_b$ are Kerr coefficients (strength of nonlinearity) for individual oscillators where $\hat{N}_a=\hat{a}^\dagger \hat{a}$ and $\hat{N}_b=\hat{b}^\dagger \hat{b}$ are number operators. $\hat{H}_{int}$ corresponds to an internal coherent coupling with coupling strength of $g$ which can be implemented in several ways \cite{Lu2023-nh,Basil,Chapman}. Finally, $\hat{H}_{drive}$ represents a resonant time-dependent drive with $\kappa_a$ and $\kappa_b$ being associated dissipation rates. This term feeds in the external input, where $\epsilon_a$ and $\epsilon_b$ are input strengths (drive amplitudes) that encode the input data $X_i$ by setting $\epsilon_a=\epsilon_b=\epsilon_0X_i$ \cite{Dudas2023-bj}.

We model the coupled Kerr oscillator system as an open quantum system with Markovian dynamics. The evolution is governed by a Lindblad master equation for the density operator $\hat{\rho}(t)$:
\begin{equation}
\frac{d\hat{\rho}(t)}{dt}= -i[\hat{H}(t),\hat{\rho}(t)]+\sum_{i=1}^4(\hat{c}_i\hat{\rho}(t)\hat{c}_i^\dagger-\frac{1}{2}\{\hat{c}_i^\dagger\hat{c}_i,\hat{\rho}(t)\})
\end{equation}
where the collapse operators describing decay in modes $a$ and $b$ are given by $\hat{c_1}=\sqrt{2\kappa_a}\hat{a}$, $\hat{c_2}=\sqrt{2\kappa_b}\hat{b}$, $\hat{c_3}=\sqrt{\kappa_\phi}\hat{N}_a$, and $\hat{c_4}=\sqrt{\kappa_\phi}\hat{N}_b$ with $\kappa_\phi$ being associated dephasing rate. Throughout this work, we set the parameters to be equal for both oscillators and, for convenience, we work in a rotating frame at the drive frequencies (taken equal to $\omega_a$ and $\omega_b$ for each mode); in this frame the free oscillation terms are eliminated (effectively setting $\omega_a = \omega_b = 0$). It is worth noting that $\hat{H}_{int}$ does not include nonlinear coupling proportional to $\hat{N}_a\hat{N}_b$, but only the linear one, described by Eq.~\ref{eq:hint}. The same Hamiltonian $\hat{H}_{nl}+\hat{H}_{int}$ was used, e.g., by Bernstein \cite{Bernstein1993-xh} and Chefles and Barnett \cite{Chefles1996-kw} to describe the nonlinear coupler. Note that we are using dimensionless units throughout this study.

\subsection{Learning and Simulation}
An input signal $s_k$ with a frequency scale $f$ is generated as follows:
\begin{equation}
 s_k=\sum_{i=1}^{20}\sin(2\pi f_i t_k +2\pi\zeta)   
\end{equation}
20 frequencies ${f_i}$ are evenly spaced in the interval $[f/5000,f/50]$. Here $t_k$ denotes the time elapsed after $k$ time steps, and $\zeta$ is a random number uniformly distributed within the interval $[0,1]$. All series are normalized to ensure a range between 0 and 1 \cite{yousef}. We use this task to have a way of tuning the level of complexity. Examples of such series at different frequency scales, denoted by $f$, are provided in section~\ref{sec:res}. 

We partition the input signal $s_k$ into data points based on the time resolution $\delta t$. As mentioned earlier, these data points ($X_i$ in Eq.~\ref{eq:hdrive}) are injected by means of a drive Hamiltonian, a widely utilized form of input encoding~\cite{PhysRevResearch.3.013077,Dudas2023-bj,Kalfus}. This strategy is particularly advantageous because it avoids the experimental overhead of state injection, allowing for highly nonlinear feature mapping~\cite{minimalhamil}. We initialize the system in the vacuum state ($\ket{0,0}$), then drive it with the first input data. The master equation describes how the state evolves under the influence of the drive and the internal Hamiltonian. We employ time multiplexing, where the reservoir state is measured in the Fock basis (readout) at intervals of $\delta t /m $, thus creating $m$ virtual nodes. These readouts occur between time steps, within a duration called the time resolution (shown by $\delta t$ in Fig.~\ref{fig:scheme}). We repeat this process for each time step until the end of the input sequence. The signals extracted from the virtual nodes, which are the expectation values of the Fock basis measurements, are used in a linear regression to predict the target: the input time series after a certain delay in the future (here, the delay is 10 across all time series used). Mathematically, the collected output vector $\mathbf{x}_k$ at time step $k$ is mapped to the target output $y_{k}$ via a learned weight matrix $W_{out}$ and a bias term $b$, such that $y_{k} = \mathbf{W}_{out} \cdot \mathbf{x}_k + b \approx s_{k+\Delta}$, where $\Delta=10$ is the prediction delay. This process is illustrated in Fig.~\ref{fig:scheme}. Throughout this work, we set $\delta t=100$ and $m=10$.

The first half of the input time series is used for training and the other half for testing.
Note that after measurements at each step, we use the final state of the system before the measurements and after the quantum evolution in the current step to continue with the next step (we do not use the collapsed state after measurements). After training is done, we re-initialize the state of the system to vacuum to start the testing process. For training, Tikhonov regularization is utilized with a regularization parameter that is chosen to get maximum performance. It is worth mentioning that both testing and training were performed in the steady state. 

For quantifying the performance of the reservoir, we use the normalized root mean square error (NRMSE), which is calculated as 
\begin{equation}
   NRMSE=\frac{1}{y_{max}-y_{min}}\sqrt{\frac{\sum_{i=1}^N(y_i-\tilde{y}_i)^2}{N}} 
\end{equation}
where $\tilde{y}$ indicates the test target and $y$ the prediction.
\subsection{Entanglement}
For quantifying the entanglement we use logarithmic negativity \cite{Vidal,Plenio}, which can be calculated as
\begin{equation}
    E_N(\rho)=\log_2||\rho^{\Gamma_A}||_1 
\end{equation}
where $\rho^{\Gamma_A}$ is the partial transpose with respect to the subsystem A. Here, $\| \cdot \|_1$ denotes the trace norm, which is defined as $\| X \|_1 = \mathrm{Tr}(\sqrt{X^\dagger X})$. Examples of the behavior of this quantity are given in section~\ref{sec:res}. for some frequency scales.
 
\subsection{\label{sec:cut}Cutoff Dimension}

To numerically simulate the coupled Kerr oscillator system, we truncate the infinite-dimensional Hilbert space by imposing an excitation number cutoff $N_{\text{cut}}$. For this study, we put $N_{\text{cut}}=3$. To ensure physical validity and convergence across different cutoffs, we constrain the mean excitation numbers $\langle \hat{n}_a \rangle$ and $\langle \hat{n}_b \rangle$ to remain below one. This ensures negligible occupation of higher Fock states to avoid truncation artifacts.

\section{\label{sec:res}Results}
\begin{figure*}[htbp]
  \centering
  \begin{tabular}{cccc}
    \includegraphics[width=0.25\textwidth]{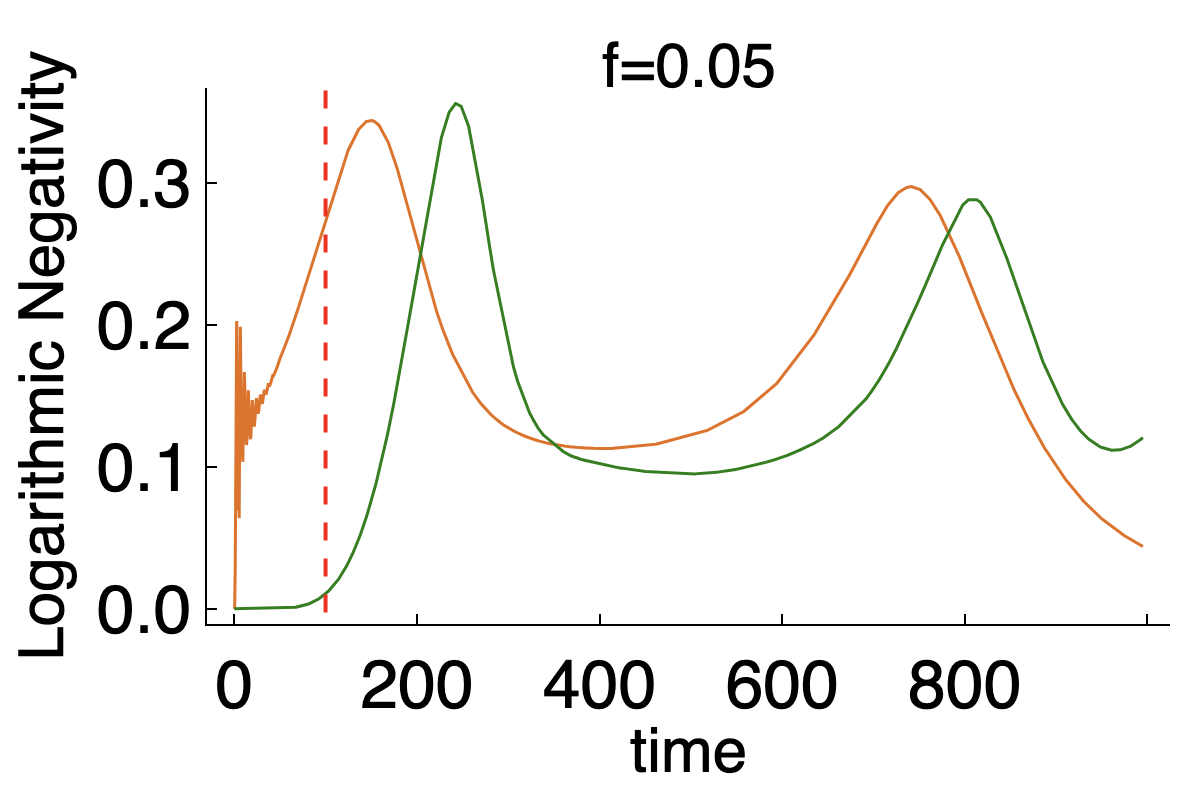} &
    \includegraphics[width=0.23\textwidth]{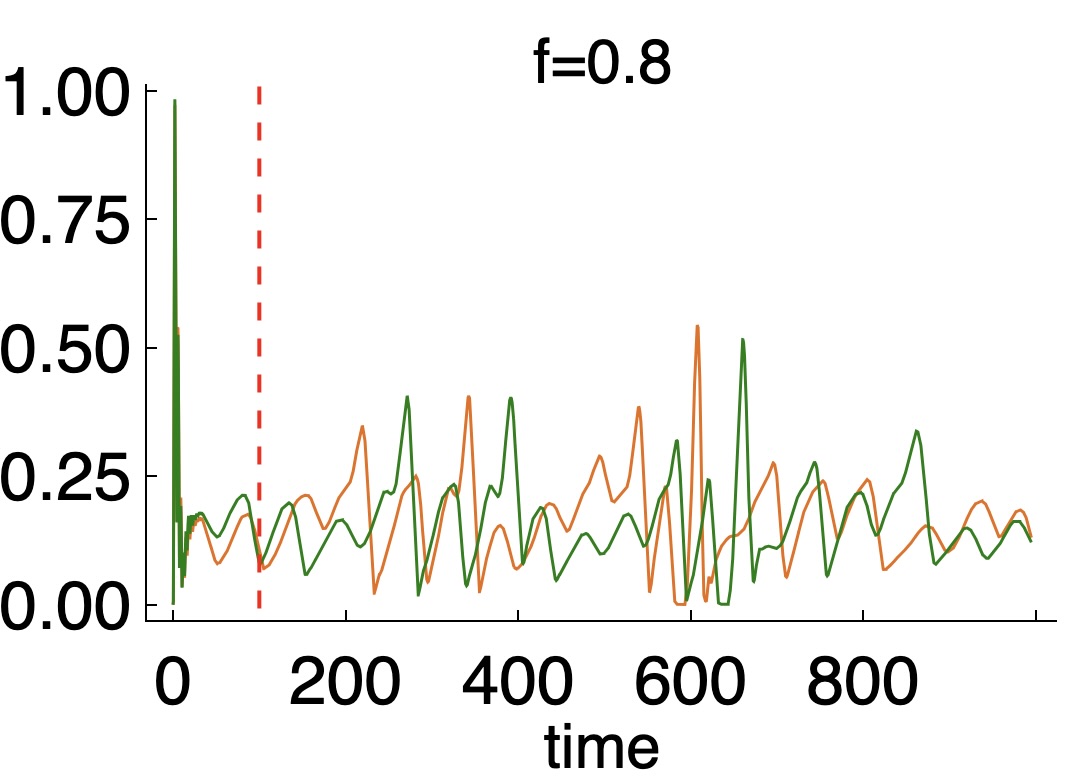} &
    \includegraphics[width=0.23\textwidth]{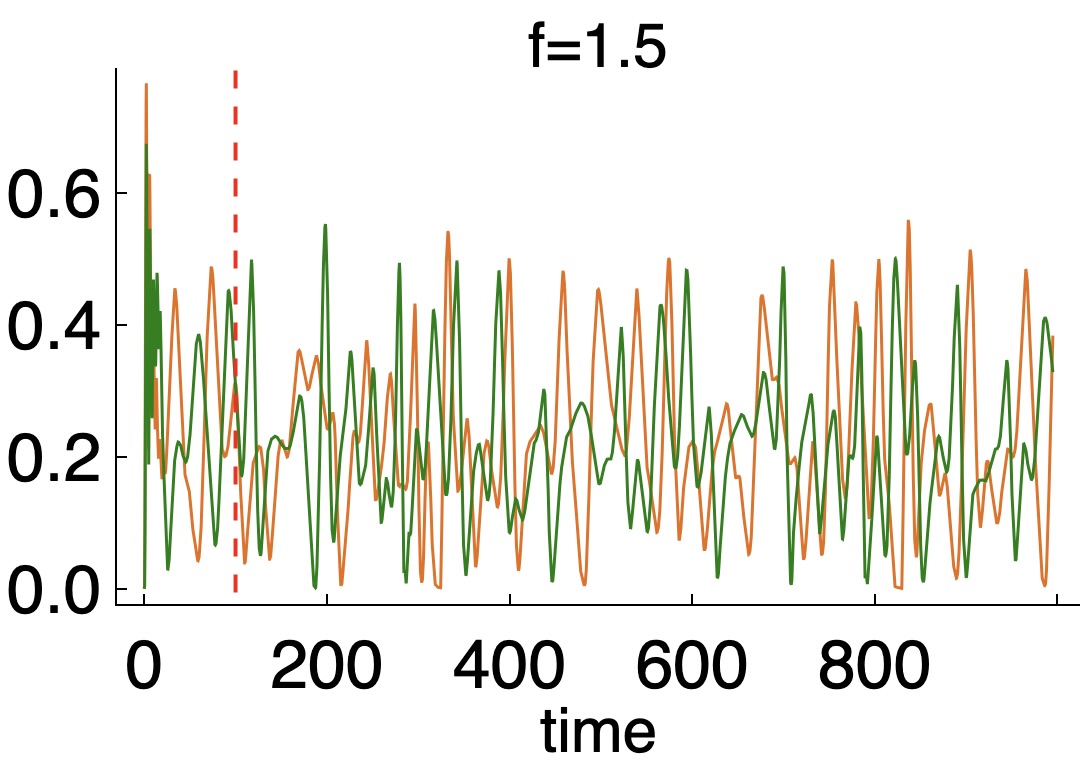} &
    \includegraphics[width=0.23\textwidth]{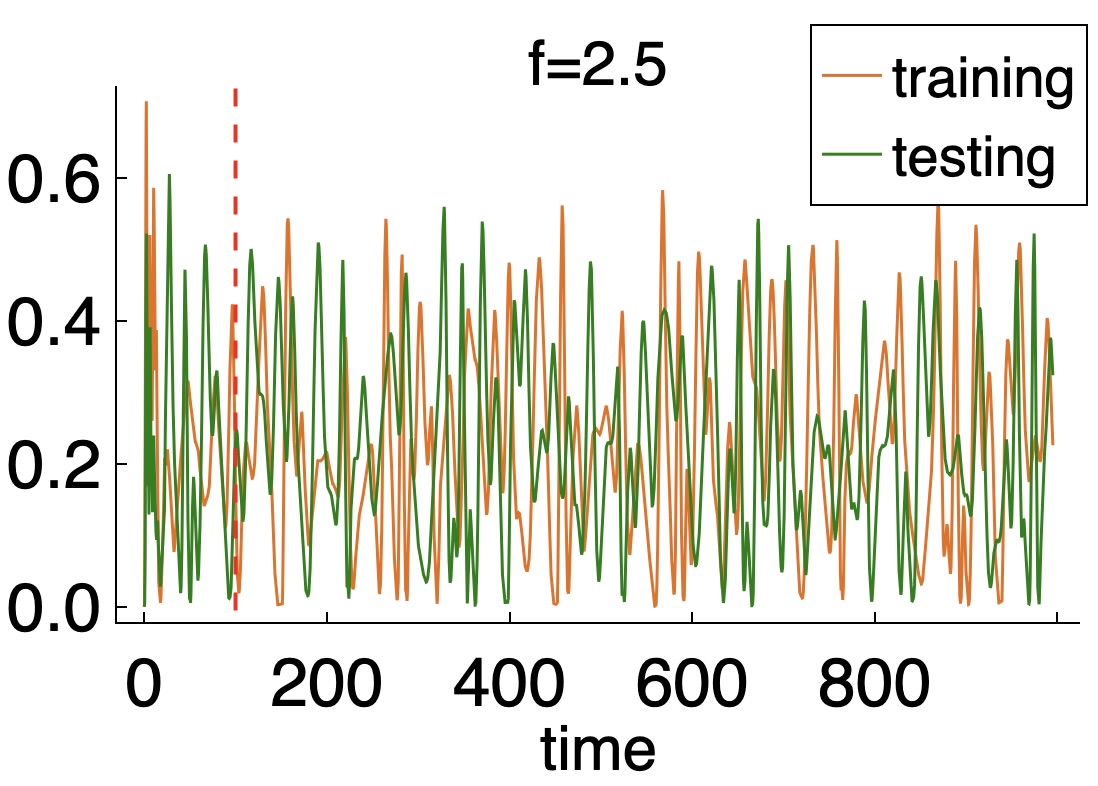} \\
    \includegraphics[width=0.25\textwidth]{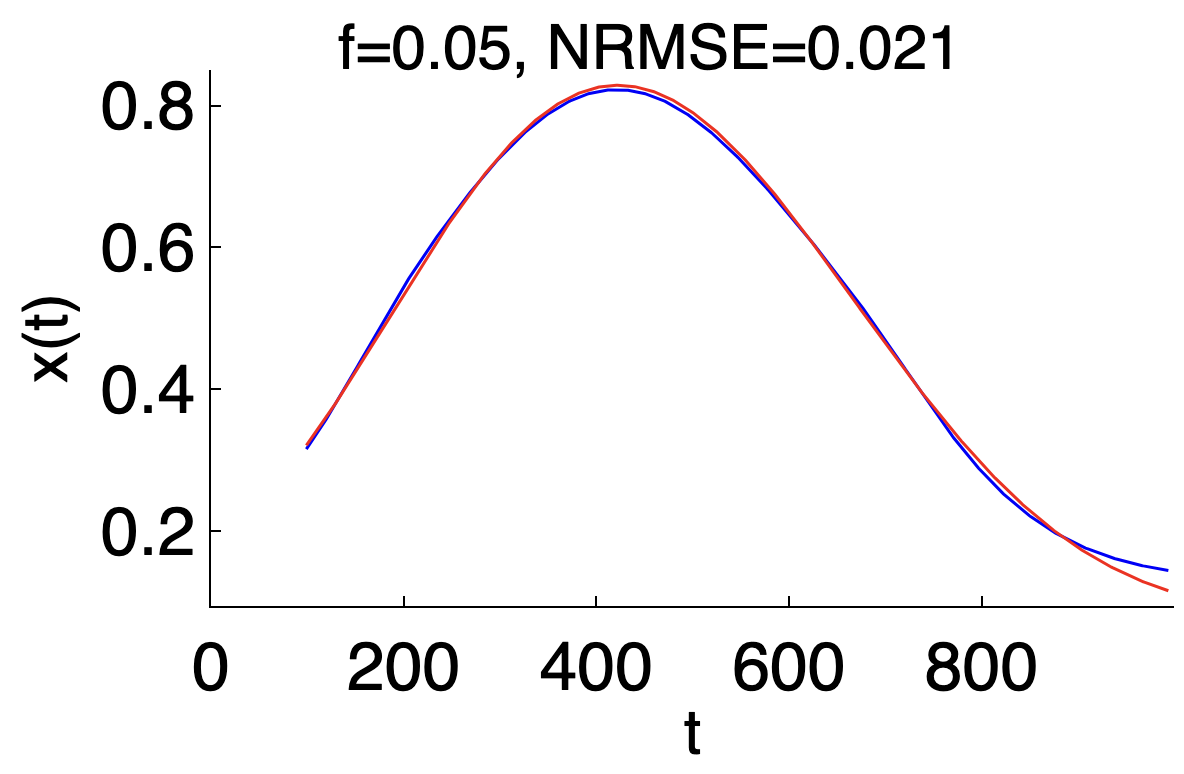} &
    \includegraphics[width=0.23\textwidth]{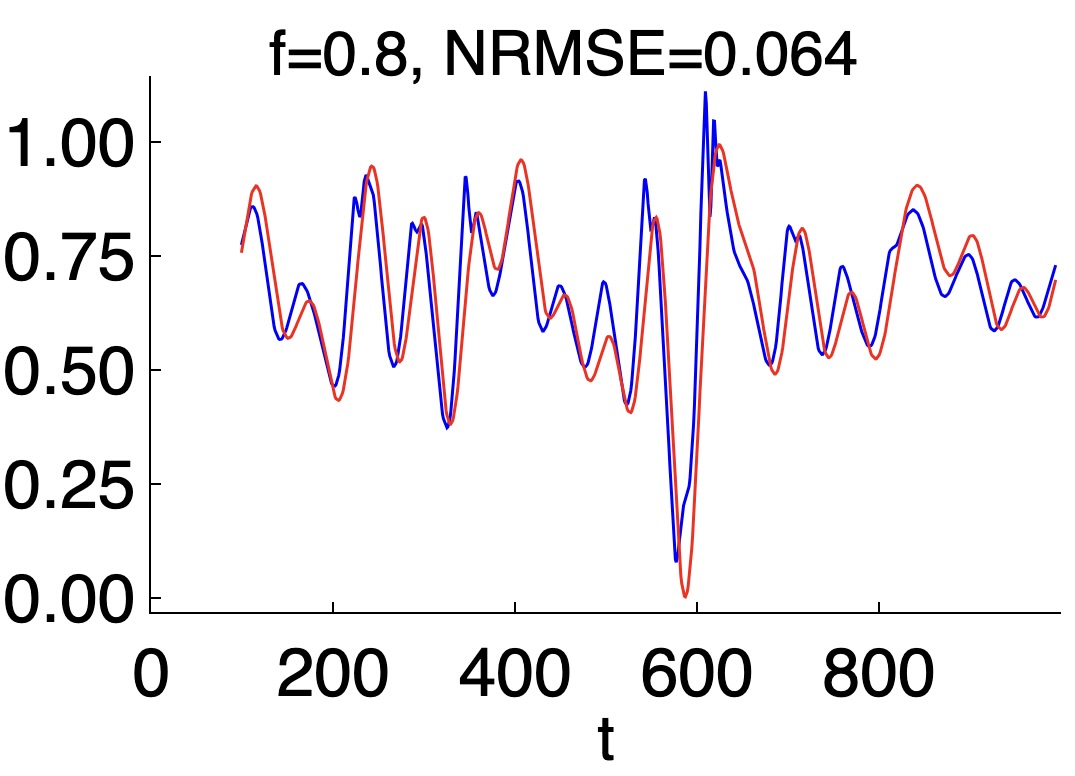} &
    \includegraphics[width=0.23\textwidth]{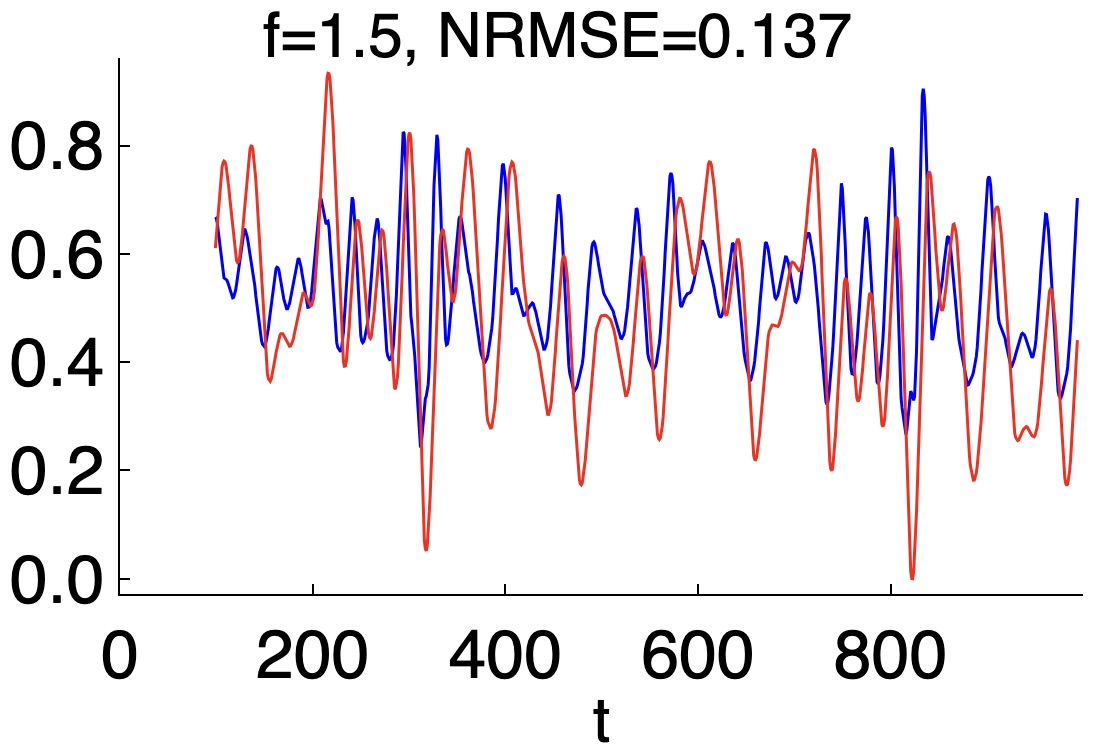} &
    \includegraphics[width=0.23\textwidth]{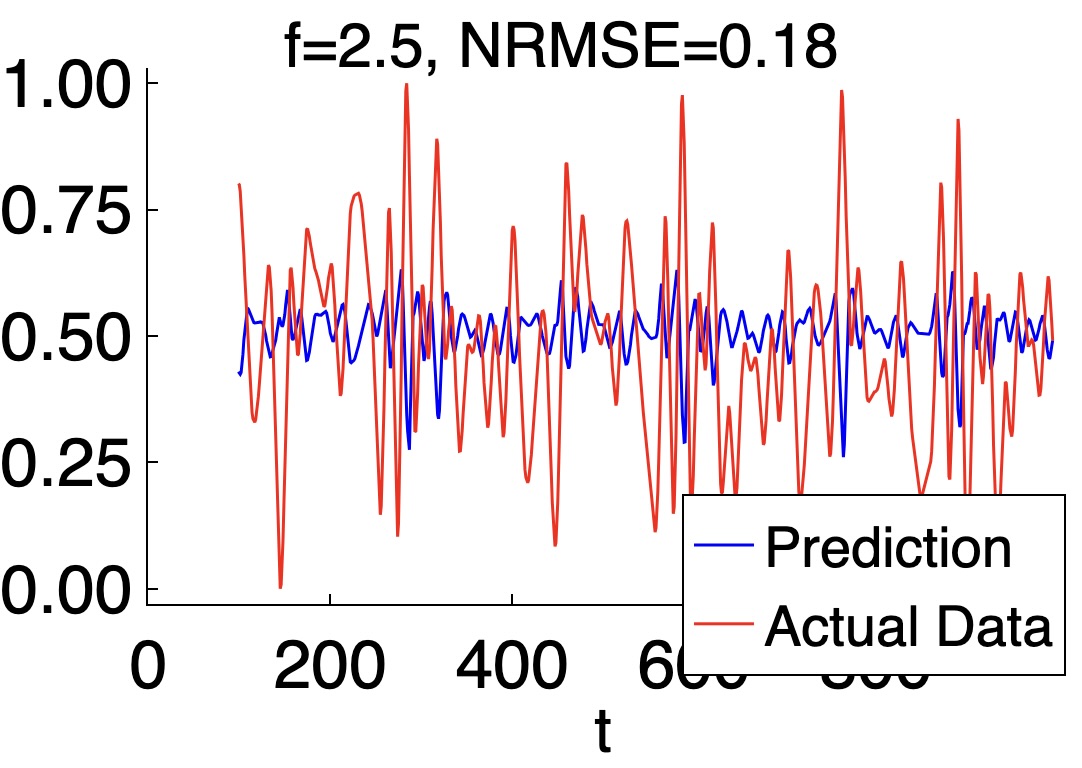} \\
  \end{tabular}
  \caption{Time evolution of entanglement (top row) and corresponding time-series prediction performance (bottom row) at different input frequencies $f$. Top: Behavior of logarithmic negativity in time during testing and training. To ensure the reservoir reflects stable dynamics rather than initial transients, we use the portion of the evolution after the vertical dashed line for training and testing. Bottom: Examples of input sequences at four different frequency scales and the corresponding predictions of the reservoir at coupling rate of $g=0.9$, input strengths of $\epsilon_a=\epsilon_b=3$, nonlinearities of $K_a=K_b=1$, and dissipation rates of $\kappa_a=\kappa_b=0.1$ in a rotating frame. The performance gets worse as the input frequency increases.}
  \label{fig:out}
\end{figure*}
In this section, we present the results of our numerical investigation into the relationship between entanglement and computational performance in a QRC framework based on coupled Kerr nonlinear oscillators. We examine the system's predictive performance on nontrivial time-series tasks, quantify entanglement using logarithmic negativity, and analyze the impact of system nonlinearity, as well as dissipation and dephasing. 

To ground the subsequent analysis, Fig.~\ref{fig:out} provides an overview of the reservoir’s behavior across different input frequencies. The top row shows the time evolution of entanglement, measured via logarithmic negativity, during training and testing.
To avoid transient effects, only the portion after a set initialization period (indicated by the dashed vertical line) is used. This removed interval (washout phase) is chosen based on the lowest dissipation rate to ensure a right choice for all the other dissipation rates of interest. The bottom row displays examples of the input time series (or target) alongside the reservoir’s predictions at selected frequencies, with fixed parameters.

The corresponding NRMSE values reveal that prediction quality declines with increasing input frequency. 

Building on these observations, we analyze the effect of each physical system parameter on performance by plotting the NRMSE against the specific parameter and simultaneously assessing the entanglement to identify potential relationships. Subsequently, we synthesize these findings through binned analysis to visualize the underlying relation more effectively.

\subsection{\label{sec:inpst}Input Strength}
\begin{figure}[htbp]
  \centering
  \begin{tabular}{ccc}
    \includegraphics[width=0.19\textwidth]{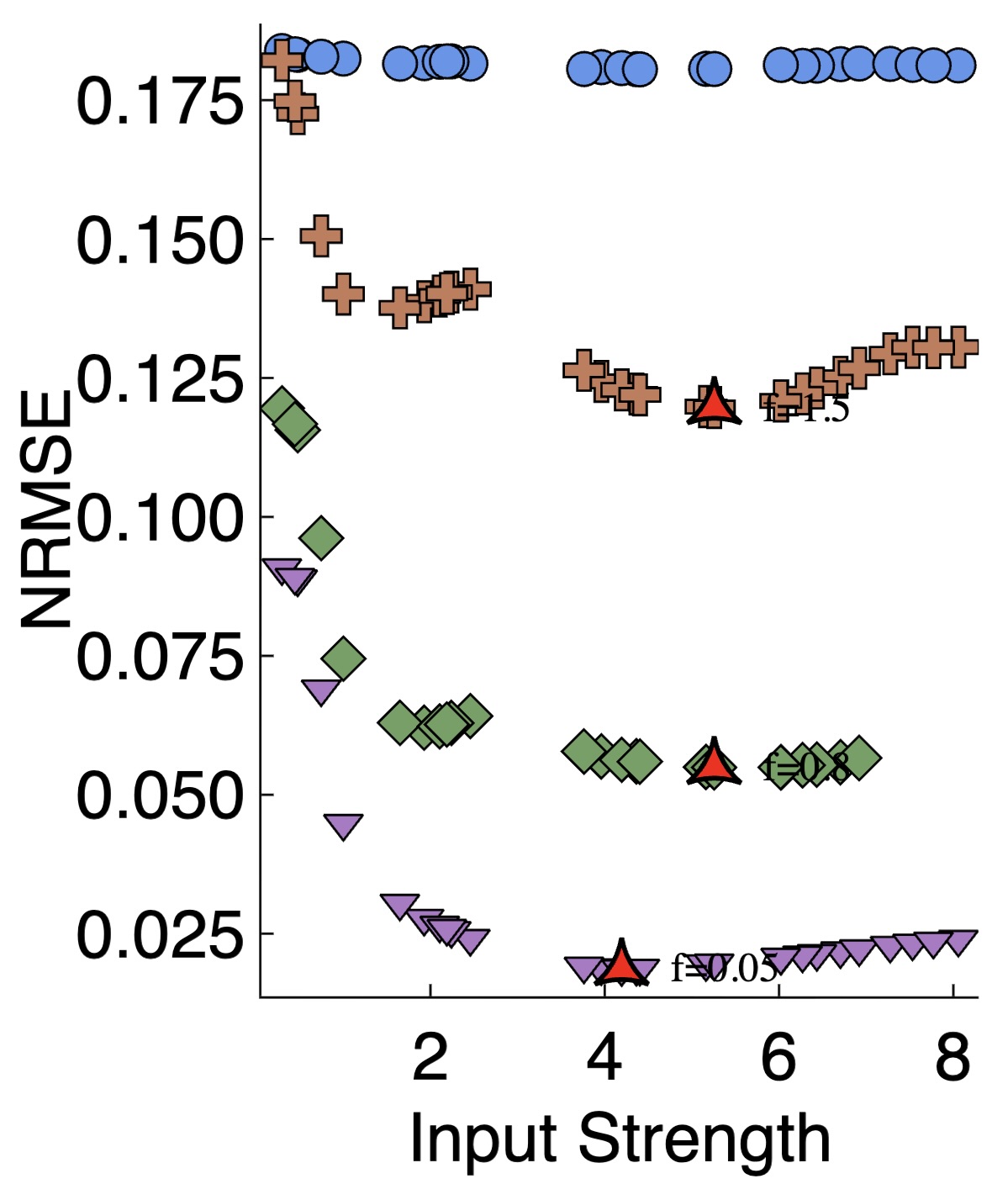} &
    \includegraphics[width=0.26\textwidth]{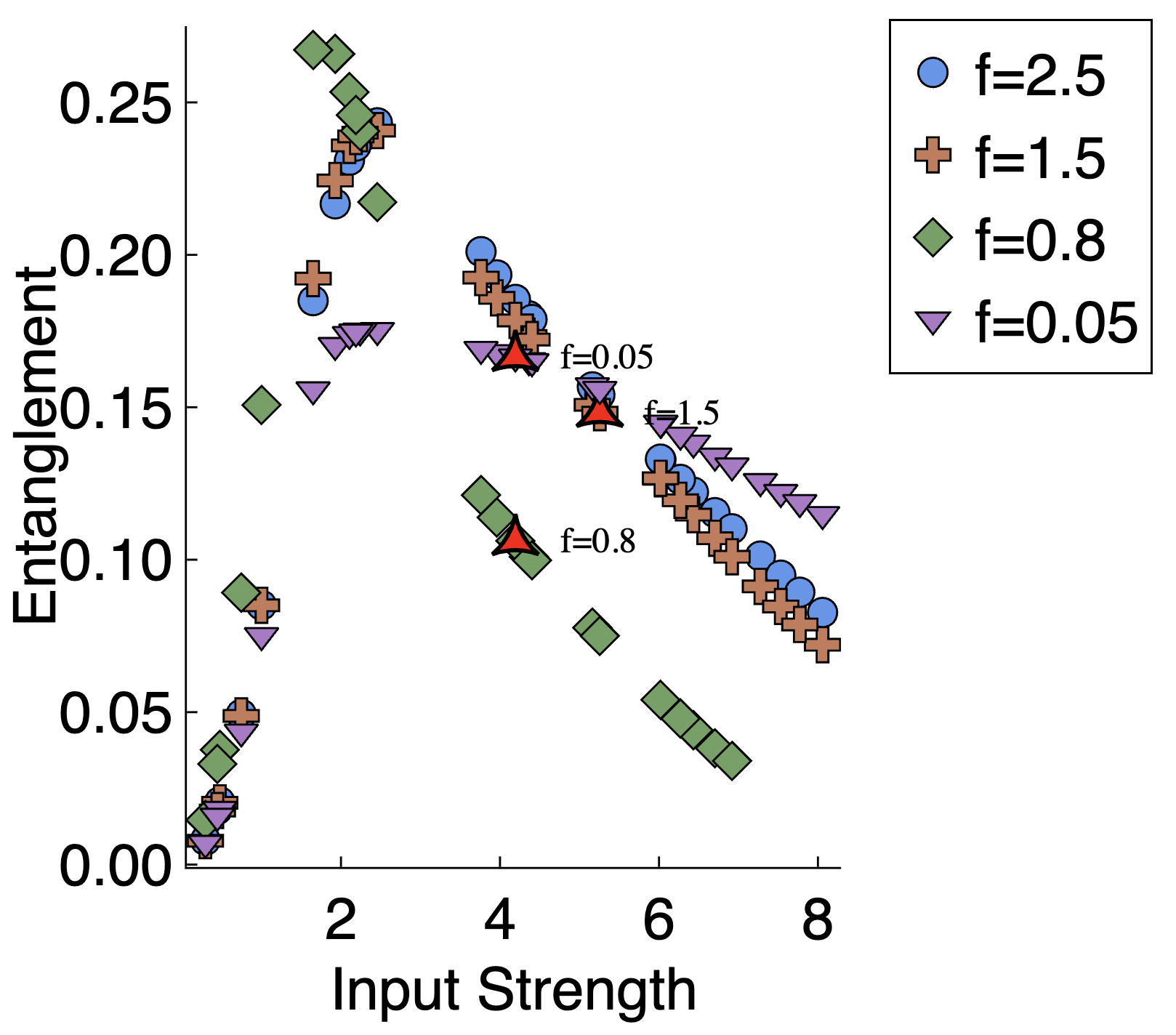} \\
  \end{tabular}
  \caption{NRMSE (left) and entanglement (right) versus input strength $\epsilon$ for different input frequencies, with fixed parameters $K=1$, $\kappa_a=\kappa_b=0.1$, $g=0.9$, and $\kappa_{\phi}=0$. Red triangles denote the specific data points that yield the absolute minimum NRMSE for each frequency. Left: Near-zero input strength yields the worst prediction error. NRMSE decreases up to an optimal basin before saturating or worsening. Right: Logarithmic negativity increases with $\epsilon$  until a peak near $\epsilon\simeq2$, beyond which it declines. By mapping the minimum NRMSE points (red triangles) to the entanglement panel, one can see that optimal forecasting performance coincides with moderate but non-zero levels of entanglement.}
  \label{fig:inpst}
\end{figure}

We randomly selected 30 input strengths ($\epsilon$) from the interval $[0,10]$, while keeping the other parameters constant, including $K_a=K_b=K=1$ and $g=0.9$, and evaluated NRMSE for input time series at various input frequency scales while computing the time average of the logarithmic negativity in the steady state of the system in each instance. It is important to note that we utilize the same 30 values of input strengths for all frequencies. Furthermore, we ensure that the average excitation number remains below one as discussed in section~\ref{sec:cut} and appendix~\ref{app:cut}. The outcome is presented in Fig.~\ref{fig:inpst}. 

Increasing input strength leads to slightly different behaviors for different frequencies. For $f=0.05$ (purple points in Fig.~\ref{fig:inpst}), NRMSE continuously decreases up to $\epsilon \simeq 4$ and then slightly increases. For $f=0.8$ (green points in Fig.~\ref{fig:inpst}), NRMSE decreases up to $\epsilon \simeq 2$, then it slightly increases and again degrades up to $\epsilon \simeq 5$, and it almost saturates afterwards. For $f=1.5$ (orange points in Fig.~\ref{fig:inpst}), a similar behavior to $f=0.8$ is observed except for after $\epsilon \simeq 5$, where NRMSE increases rather than saturating. Finally, for $f=2.5$ (orange points in Fig.~\ref{fig:inpst}), no improvement is observed. This is because the task becomes too difficult for our system, which has a limited cutoff dimension and average excitation number, a phenomenon that might not occur in a practical setting.

By focusing on the entanglement (right panel of Fig.~\ref{fig:inpst}), we observe that it increases up to $\epsilon \simeq 2$ for all frequencies. Beyond this point, entanglement decreases, leading to a peak in entanglement. Crucially, near-zero input strength yields the absolute highest prediction errors across all frequencies. Tracing the absolute minimum NRMSE for each frequency (marked by red triangles) reveals that the optimal predictive basin coincides with robust, moderate but non-zero entanglement levels (e.g., ranging between $0.11$ and $0.17$ for $f \leq 1.5$). Therefore, while the best performance is not achieved with the maximum entanglement, it coincides with a moderate level of entanglement.

\subsection{\label{sec:nl}Kerr Nonlinearity}

\begin{figure}[htbp]
  \centering
  \begin{tabular}{ccc}
    \includegraphics[width=0.19\textwidth]{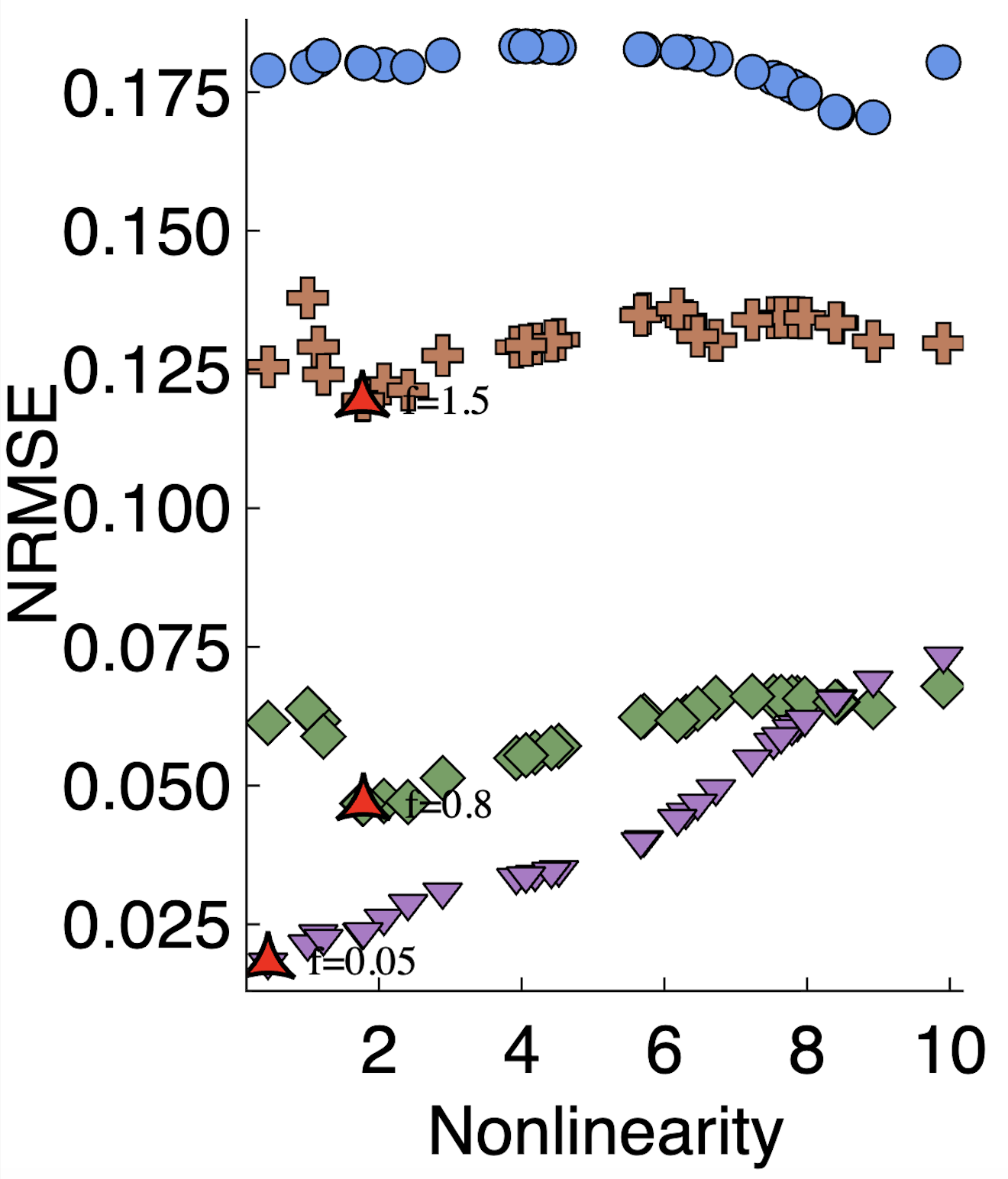} &
    \includegraphics[width=0.26\textwidth]{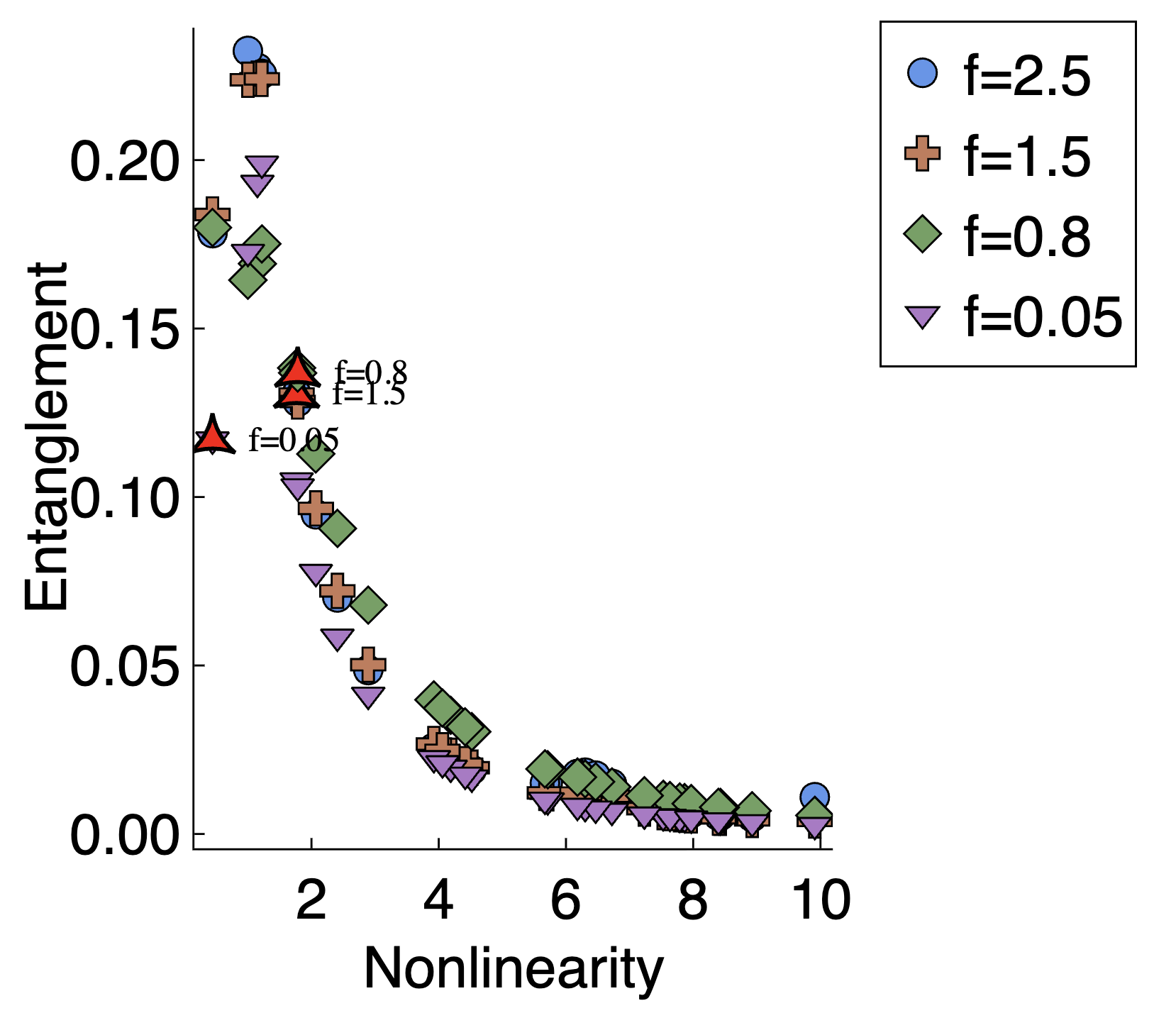} \\
  \end{tabular}
  \caption{NRMSE (left) and Entanglement (right) versus Kerr nonlinearity $K$ for different input frequencies, with fixed parameters $\epsilon=3$, $\kappa_a=\kappa_b=0.1$, $g=0.9$, and $\kappa_\phi=0$. Red triangles mark the data points corresponding to the absolute minimum NRMSE for each frequency below the threshold. Left: NRMSE fluctuates up to $K\simeq2$ for $f=0.8$ and $f=1.5$ and increases thereafter. Right: Entanglement fluctuates up to $K\simeq1$, beyond which it decreases. Tracing the optimal performance points (red triangles) confirms that the minimum prediction errors coincide with a moderate but non-zero level of entanglement.}
  \label{fig:knl}
\end{figure}

We randomly selected 30 Kerr nonlinearity values ($K$) from the interval $[0,10]$, while keeping the other parameters constant, including $\epsilon_a=\epsilon_b= \epsilon= 3$ and $g=0.9$, and evaluated the NRMSE for input time series at different input frequency scales while calculating the time average of the logarithmic negativity of the system in each case. Note that we use the same 30 values of $K$ for all frequencies. The results are shown in Fig.~\ref{fig:knl}.

The NRMSE for $f=0.8$ and $f=1.5$ (green and orange points in Fig.~\ref{fig:knl}) fluctuates, reaching minimum values near $K \simeq2$ before increasing. For $f=0.05$, the minimum error occurs at very low nonlinearity, after which it increases monotonically, and for $f=2.5$, no clear trend is observed. By explicitly mapping the global minimum NRMSE for each frequency onto the entanglement plot (indicated by red triangles), we observe a consistent pattern. The best-case prediction errors for $f=0.05$, $f=0.8$, and $f=1.5$ do not occur in the zero-entanglement regime. Instead, these optimal points coincide tightly with a moderate entanglement range spanning from approximately $0.12$ to $0.14$. Beyond this optimal region ($K\simeq 2$), entanglement generally decreases while the prediction error increases.

\begin{figure}[htbp]
  \centering
  \begin{tabular}{ccc}
    \includegraphics[width=0.19\textwidth]{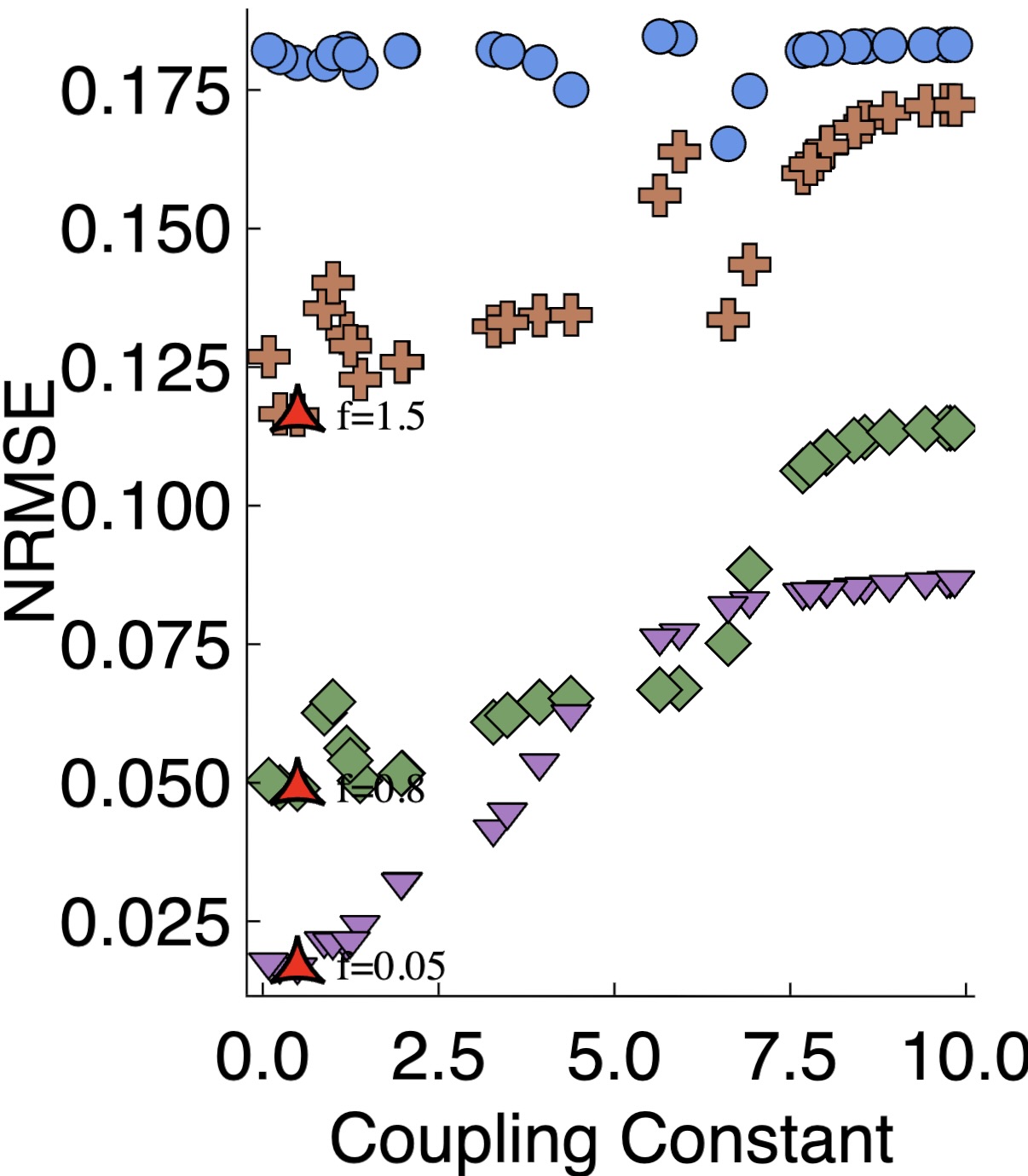} &
    \includegraphics[width=0.26\textwidth]{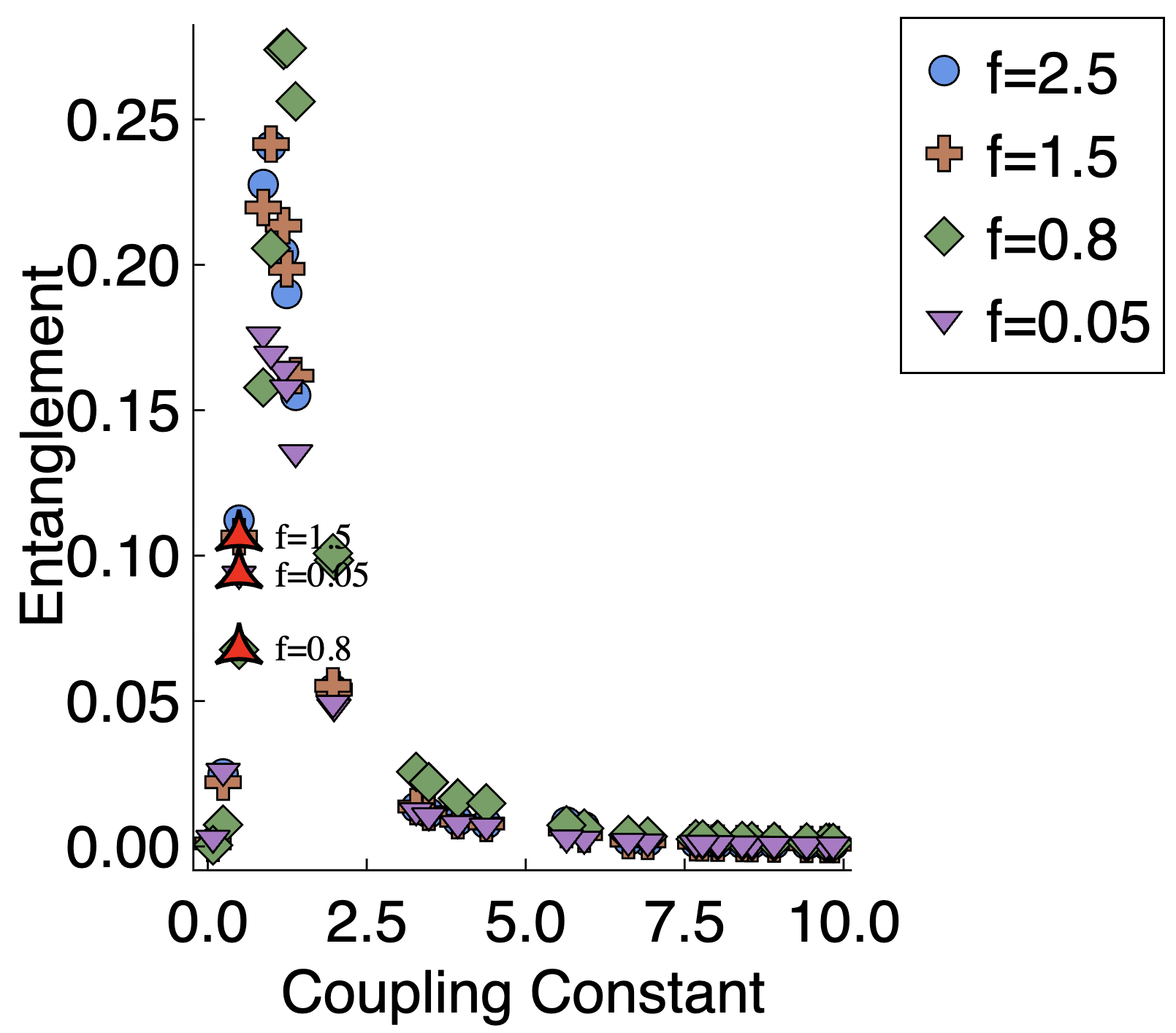} \\
  \end{tabular}
  \caption{NRMSE (left) and Entanglement (right) versus coupling constant $g$ for different input frequencies, with fixed parameters $\epsilon=3$, $\kappa_a=\kappa_b=0.1$, $K=1$, and $\kappa_\phi=0$. Red triangles highlight the specific parameter instances yielding the lowest NRMSE for each valid frequency. Left: NRMSE fluctuates before eventually increasing at higher coupling strengths. Right: Entanglement peaks near $g \simeq 0.9$, beyond which it decreases. Notably, while the optimal coupling strength is small, the corresponding entanglement is non-zero. One can see in this figure that, once again, optimal forecasting performance does not occur in the zero-entanglement state, aligning instead with a moderate entanglement regime.}
  \label{fig:cc}
\end{figure}

\subsection{\label{sec:cc}Coupling Constant}

We randomly selected 30 coupling constant values ($g$) from the interval $[0,10]$, while keeping the other parameters constant, including $K_a=K_b=K=1$ and $\epsilon_a=\epsilon_b= \epsilon= 3$, and evaluated the NRMSE on input time series at different input frequency scales while calculating time average of the logarithmic negativity of the system in each case. Note that we use the same 30 values of $g$ for all frequencies. The results are shown in Fig.~\ref{fig:cc}.

The relation is very similar to the case of nonlinearity. The NRMSE for $f = 0.8$ and $f = 1.5$ (green and orange points in Fig.~\ref{fig:cc}) fluctuates up to $g \simeq 1.4$. Beyond this point, The NRMSE starts increasing. For $f=0.05$, it increases monotonically, and for $f = 2.5$, no clear trend is observed (left panel of Fig.\ref{fig:cc}).

By focusing on the entanglement (right panel of Fig.~\ref{fig:cc}), we observe that it peaks at $g \simeq 0.9$. By examining the regime of minimum prediction error (indicated by the red triangles in Fig.~\ref{fig:cc}), We observe that while intermediate frequencies achieve minimum error at weak coupling ($g\simeq0.1$), the right panel demonstrates that the entanglement at these points is not zero, but has risen to moderate but non-zero levels ($E\simeq0.06$ to $0.11$). Across all valid frequencies, the best-case error coincides with the moderate but non-zero entanglement regime.

In summary, across the preceding parameter sweeps, we consistently observe that the lowest prediction error (NRMSE) occurs at moderate but strictly non-zero entanglement, suggesting a positive role for entanglement in these predictive tasks.

\subsection{Overall Entanglement-Performance Relationship}

To complement the preceding single-parameter sweeps, which focused on identifying the regimes that yield optimal performance, we now analyze the average predictive performance across the combined data. We perform a binned analysis by grouping the data from sections \ref{sec:inpst}, \ref{sec:nl}, and \ref{sec:cc} into equal-sized bins based on their entanglement levels and computing the mean NRMSE for each. This complementary approach allows us to observe the broader, overall relationship between entanglement and the reservoir's forecasting capability.

\begin{figure}[htbp]
  \centering
    \includegraphics[width=0.45\textwidth]{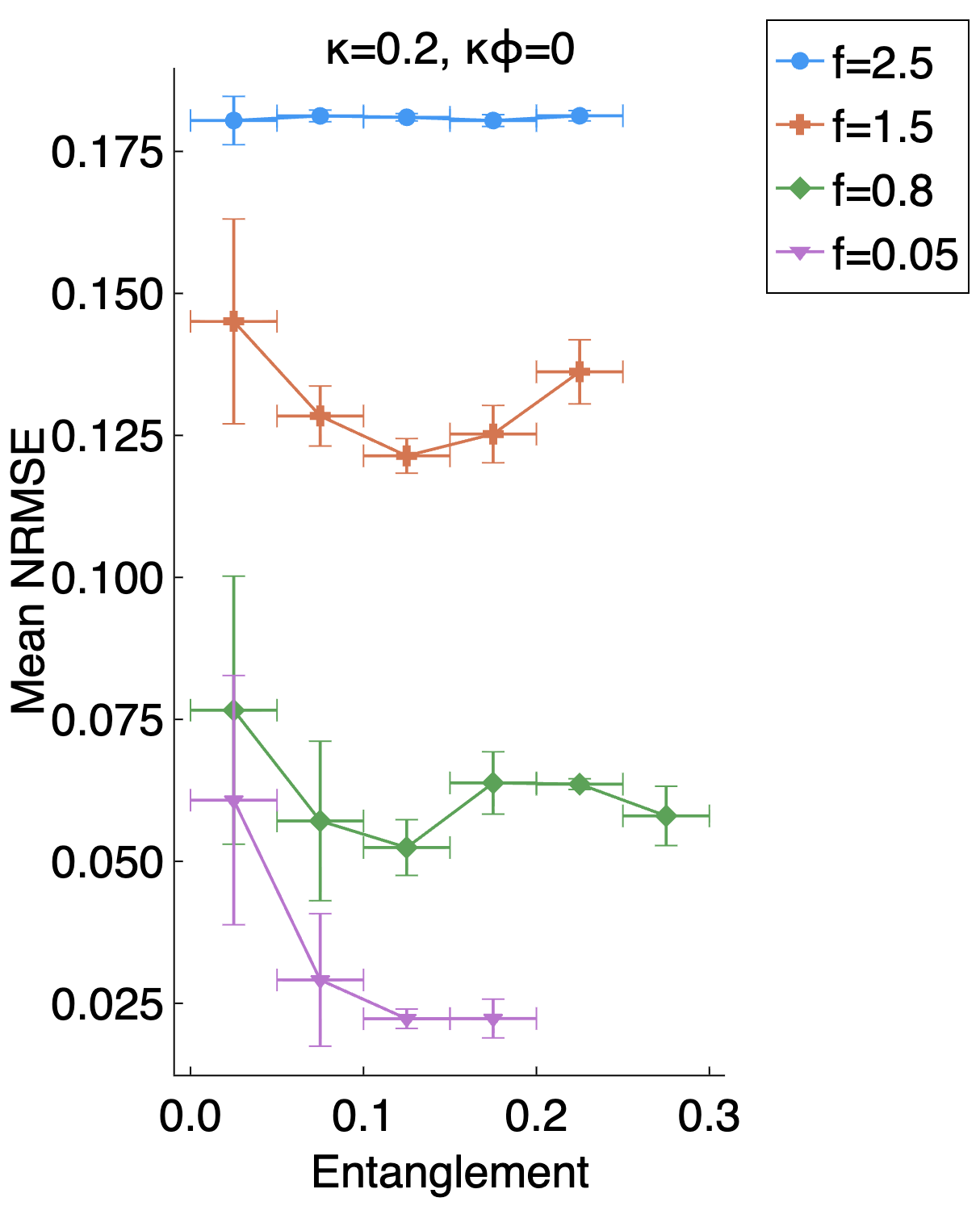}
  \caption{Prediction error (NRMSE) versus entanglement across different input frequencies, with} $\kappa_\phi = 0$ and $\kappa_a = \kappa_b = 0.1$. The data from \cref{fig:inpst,fig:knl,fig:cc} are combined and binned according to entanglement levels (quantified via logarithmic negativity). The plotted mean NRMSE for each bin captures the overall relationship between entanglement and performance, where vertical error bars denote the intra-bin standard deviation. The results demonstrate that a moderate but non-zero level of entanglement improves prediction accuracy. However, this trend vanishes above a frequency threshold($f=2.5$), beyond which entanglement yields no observable advantage.
  \label{fig:ent}
\end{figure}

Our analysis reveals a non-monotonic relationship between entanglement and predictive performance. As shown in Fig.~\ref{fig:ent},
we observe that moderate but non-zero levels of entanglement are associated with improved average predictive performance, with NRMSE reaching a minimum, within a specific frequency regime of the input. However, for frequency scale $f=0.05 $ (purple curve in Fig.~\ref{fig:ent}), error saturates for higher entanglement, whereas a dip is observed for other frequencies below $f=2.5$. Beyond this regime, increasing the input frequency leads to a breakdown in reservoir memory and a decline in average performance, despite sustained or even increasing levels of entanglement, as the task becomes too difficult for the predictor in this setup. A potential reason for this could be the low excitation number limit that was implemented to reduce the artifacts caused by the cutoff dimension (\ref{sec:cut} and appendix~\ref{app:cut}).

\subsection{Role of Dissipation}
\begin{figure}[htbp]
  \centering
    \includegraphics[width=0.42\textwidth]{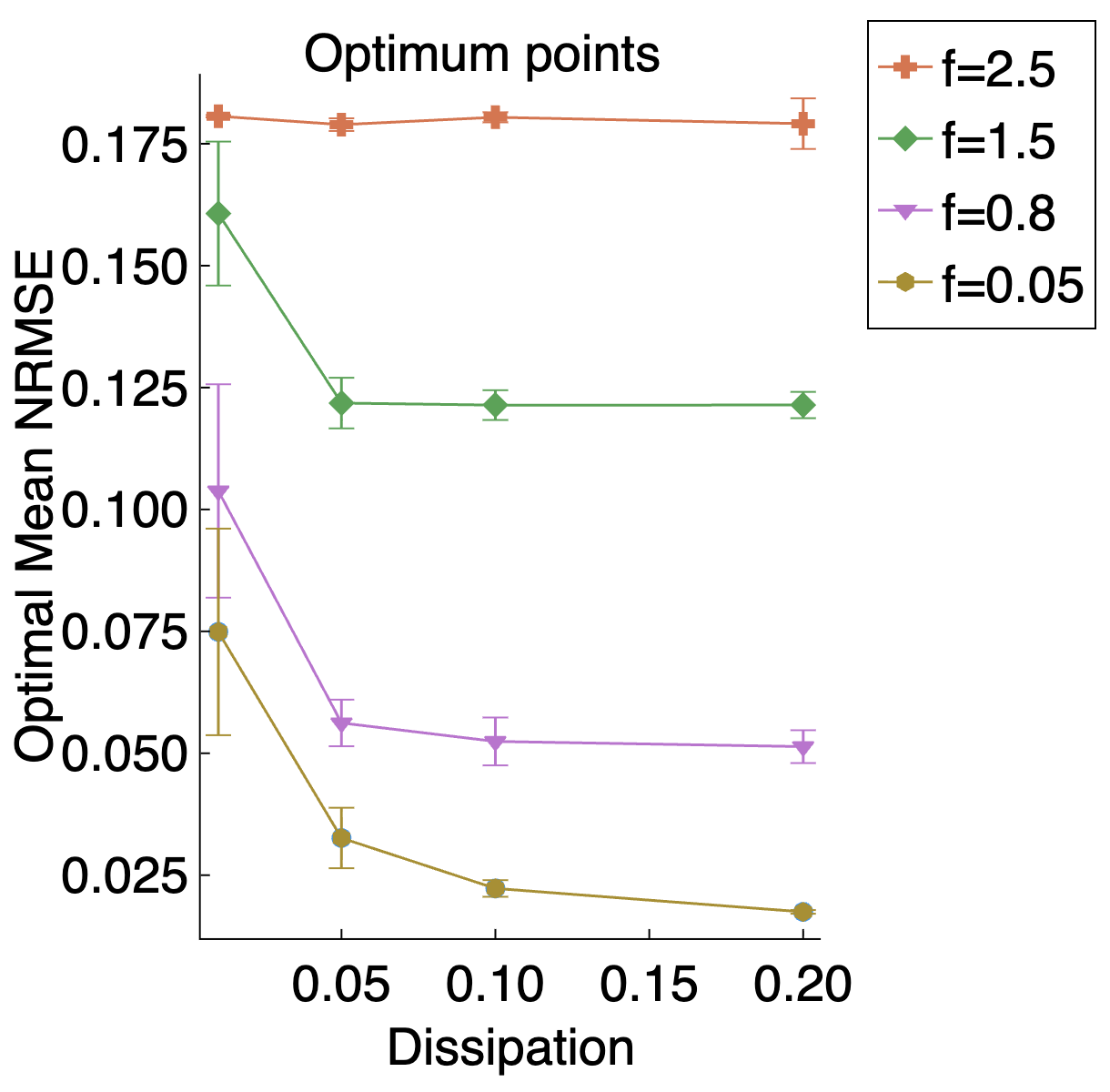} 
  \caption{Optimum mean NRMSE versus dissipation rate for different input frequencies, with $\kappa_\phi=0$. These data points correspond to the optimum values extracted from the mean NRMSE traces in Fig.~\ref{fig:ent} and Fig.~\ref{fig:pk} in Appendix~\ref{sec:apa} (error bars show the same standard deviations as Fig.}~\ref{fig:ent} and Fig.~\ref{fig:pk}). Higher dissipation improves performance for all frequencies below the frequency threshold, up to a point in dissipation where it tends to saturate.
  \label{fig:diss}
\end{figure}

 To examine the impact of dissipation, we repeat the data generation process and binned analysis described earlier in the presence of different dissipation rates, $\kappa_a=\kappa_b=\kappa$ and compute the average entanglement and NRMSE to produce the same type of plot as in Fig.~\ref{fig:ent}, now for different $\kappa$ (see Fig.~\ref{fig:pk} in Appendix~\ref{sec:apa}). It is worth mentioning that our system does not reach a steady state in the absence of dissipation and therefore does not exhibit steady-state entanglement; hence, we do not consider the zero-dissipation case. To better analyze the effect of dissipation, we extract the optimal points (those with minimum error) and plot their corresponding NRMSE as a function of dissipation for different input frequencies, resulting in Fig.~\ref{fig:diss}.
 
As shown in Fig.~\ref{fig:diss}, a higher dissipation level is better for frequencies below the frequency threshold up to a point where it saturates. By looking at Fig.~\ref{fig:pk}, Appendix~\ref{sec:apa}, one can see that observing an entanglement-performance relationship requires sufficient dissipation, with slightly higher values offering a consistent advantage across all frequency scales below the frequency threshold.
\subsection{Role of Dephasing}
To examine the impact of dephasing, we repeat the data generation process and binned analysis described earlier in the presence of different dephasing rates $\kappa_\phi$, and compute the average entanglement and NRMSE to produce the same type of plot as in Fig.~\ref{fig:ent}, now for different $\kappa_\phi$ (Fig.~\ref{fig:pd} in Appendix~\ref{sec:apa}). To more clearly assess the role of dephasing, we identify the optimal points—those corresponding to the minimum prediction error—and plot their associated NRMSE as a function of dephasing for different input frequencies, as shown in Fig.~\ref{fig:deph}.

As one can see in Fig.~\ref{fig:deph} and Fig.~\ref{fig:pd} in Appendix~\ref{sec:apa}, zero dephasing is the optimal regime both for overall performance and maintaining the entanglement-performance relationship (because at high dephasing we do not have much entanglement). However, focusing on Fig.~\ref{fig:deph} reveals that for very low frequency (e.g. $f=0.05$), some amount of dephasing could be beneficial.
\begin{figure}[htbp]
  \centering
  \begin{tabular}{ccc}
    \includegraphics[width=0.42\textwidth]{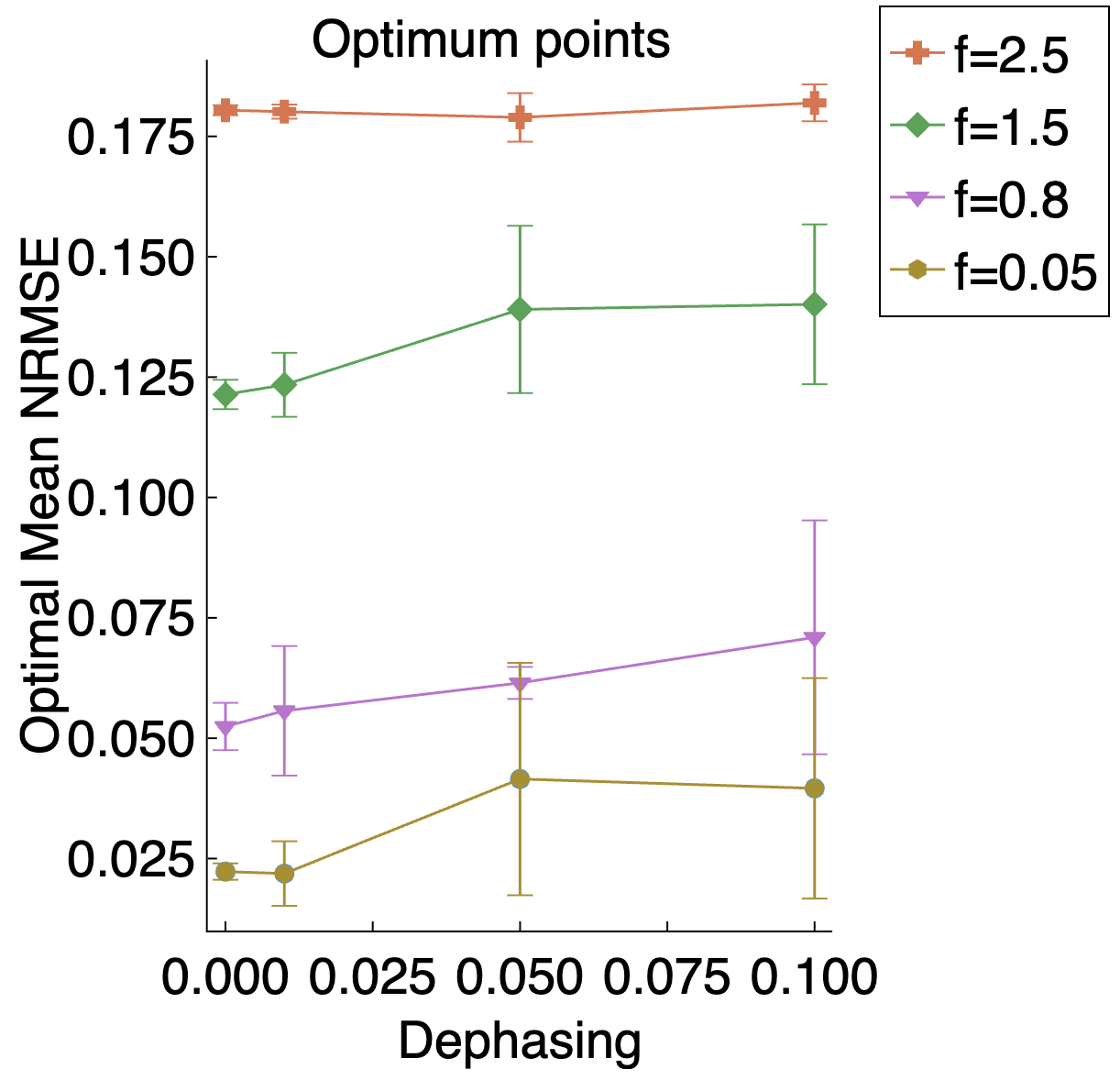} 
  \end{tabular}
  \caption{Optimum mean NRMSE versus dephasing rate for different input frequencies, with $\kappa_a=\kappa_b=0.1$. these data points correspond to the optimum values extracted from the mean NRMSE traces in Fig.}~\ref{fig:ent} and Fig.~\ref{fig:pd} in Appendix~\ref{sec:apa} (error bars show the same standard deviations as Fig.~\ref{fig:ent} and Fig.~\ref{fig:pd}). Zero or very small dephasing yields the best performance in most frequency regimes. However, at very low frequency ($f=0.05$), a small amount of dephasing seems to slightly improve prediction performance.
  \label{fig:deph}
\end{figure}

\subsection{Nonlinear Prediction Task}
To verify the generality of our findings beyond the linear forecasting task, we extended our analysis to a more complex nonlinear prediction task. We applied the 20th-order NARMA kernel to the multi-frequency input signal to generate a highly nonlinear target trajectory. As detailed in Appendix~\ref{app:narma}, our analysis confirms that the key trends observed in the linear task remain robust. Specifically, we find that moderate entanglement continues to be linked to the optimal predictive performance. Furthermore, the frequency threshold effect persists, where the relation between entanglement and performance is lost at higher input frequencies. While dissipation remains beneficial up to a saturation point, we observe that dephasing is strictly detrimental to performance in this nonlinear regime, highlighting the increased sensitivity of nonlinear tasks to phase coherence. Full details and parameter sweeps for this task are provided in Appendix~\ref{app:narma}.

\section{Discussion and Conclusion}
In this work, we presented a systematic analysis of how the computational performance of a quantum reservoir composed of two coupled Kerr nonlinear oscillators depends on various physical parameters, including input drive strength, Kerr nonlinearity, coupling strength, dissipation, and dephasing. We identified optimum performance regimes for the system's parameters. Our goal was to investigate whether and under what conditions quantum entanglement within the reservoir contributes to an improvement in time-series prediction tasks.

Our findings indicate that within the various system configurations analyzed, moderate but non-zero entanglement coincides with the best predictive performance, albeit only within a certain range of dephasing rate and up to a threshold in the input frequency. It was noted that optimal average performance, measured in terms of NRMSE, generally takes place when the reservoir maintains a moderate level of entanglement rather than the highest, with the worst-case performance following a similar trend. This effect was most evident for lower input frequency scales, where the temporal structure of the input could be effectively retained and processed by the reservoir dynamics. This kind of relation between quantumness and learning capabilities is consistent with other recent findings in the literature~\cite{motamedi2024,G_tting_2023,yousef}, which suggest that excessive “quantumness” may saturate the reservoir or introduce dynamical instability, ultimately hampering learning. However, unlike the spin-network results in \cite{yousef} which observed an increasing advantage at high frequencies, our coupled-oscillator system exhibits a distinct frequency threshold, which might be related to the excitation number cut-off and the corresponding limit on the average excitation number that we imposed.

Notably, our results demonstrate that the observed relationship between moderate entanglement and performance persists under realistic conditions with dissipation and dephasing (please see Appendix~\ref{sec:apa}). We showed that increasing dissipation can actually improve performance, up to a point that it saturates. Likewise, we found that the absence of dephasing is generally optimal for performance and entanglement (please see Appendix~\ref{sec:apa}); a small amount of dephasing can be tolerated (and at very low input frequency, it even provided a slight benefit).

There are several limitations to note. First, our study was limited to a two-mode Kerr system with a truncated Hilbert space—a limitation that we addressed by imposing a limit on average excitation number (please see Section~\ref{sec:cut}, Appendix~\ref{app:cut}). While this setup is well-suited for exploring foundational aspects of QRC, it restricts both the memory depth and the expressive capacity of the reservoir. Second, we focused on a specific class of input signals and employed a standard linear readout with ridge regression. Although this choice facilitates controlled benchmarking, our conclusions may not directly generalize to more diverse or noisy targets or to different output architectures. Additionally, our quantification of entanglement relied on logarithmic negativity, which does not capture all forms of non-classical correlation, such as Wigner negativity, quantum discord, magic/nonstabilizerness, or coherence.

While our results exhibit a consistent relationship between entanglement and prediction accuracy, we note that establishing strict causality in physical reservoirs is challenging. Because our methodology relies on one-dimensional parameter sweeps and binned statistical analysis, we must be cautious in claiming a definitive causal “advantage”. However, the robustness of this relation—persisting across different parameters, noise levels, and tasks (including the nonlinear NARMA time-series prediction task, as shown in Appendix \ref{app:narma})—suggests that moderate entanglement may play a positive role in the reservoir's forecasting power in the quantum regime.
Moreover, this finding lends some support to the broader hypothesis that quantum resources can be beneficial in machine learning frameworks, provided that system parameters and noise levels are appropriately balanced. Looking forward, extending this analysis to larger oscillator networks would allow for exploration of more complex spatial and temporal entanglement structures. It would also be valuable to consider alternative quantum resource measures and output processing techniques, including non-linear readout layers or adaptive hybrid quantum-classical schemes.
\section*{Code Availability}
The numerical code used to generate the results in this paper is available on \href{https://github.com/alikauc/Kerr_Coupled_QRC-Paper-}{GitHub}. Supplementary scripts and datasets can be obtained from the corresponding author upon reasonable request.
\acknowledgments       
This work was supported by the National Research Council
through its Applied Quantum Computing Challenge Program, the Natural Sciences and Engineering Research Council
(NSERC) of Canada through its NSERC Discovery Grant
Program, the Alberta Major Innovation Fund, and Quantum
City.  

\appendix
\section{Effects of Dissipation and Dephasing on the Overall Entanglement-Performance Relationship}
\label{sec:apa}
To further illuminate the relationship between entanglement and performance discussed in the main text, we provide additional figures and analysis.
\begin{figure*}[htbp]
  \centering
  \begin{tabular}{ccc}
    \includegraphics[width=0.32\textwidth,trim={0 0 {.28\textwidth} 0},clip]{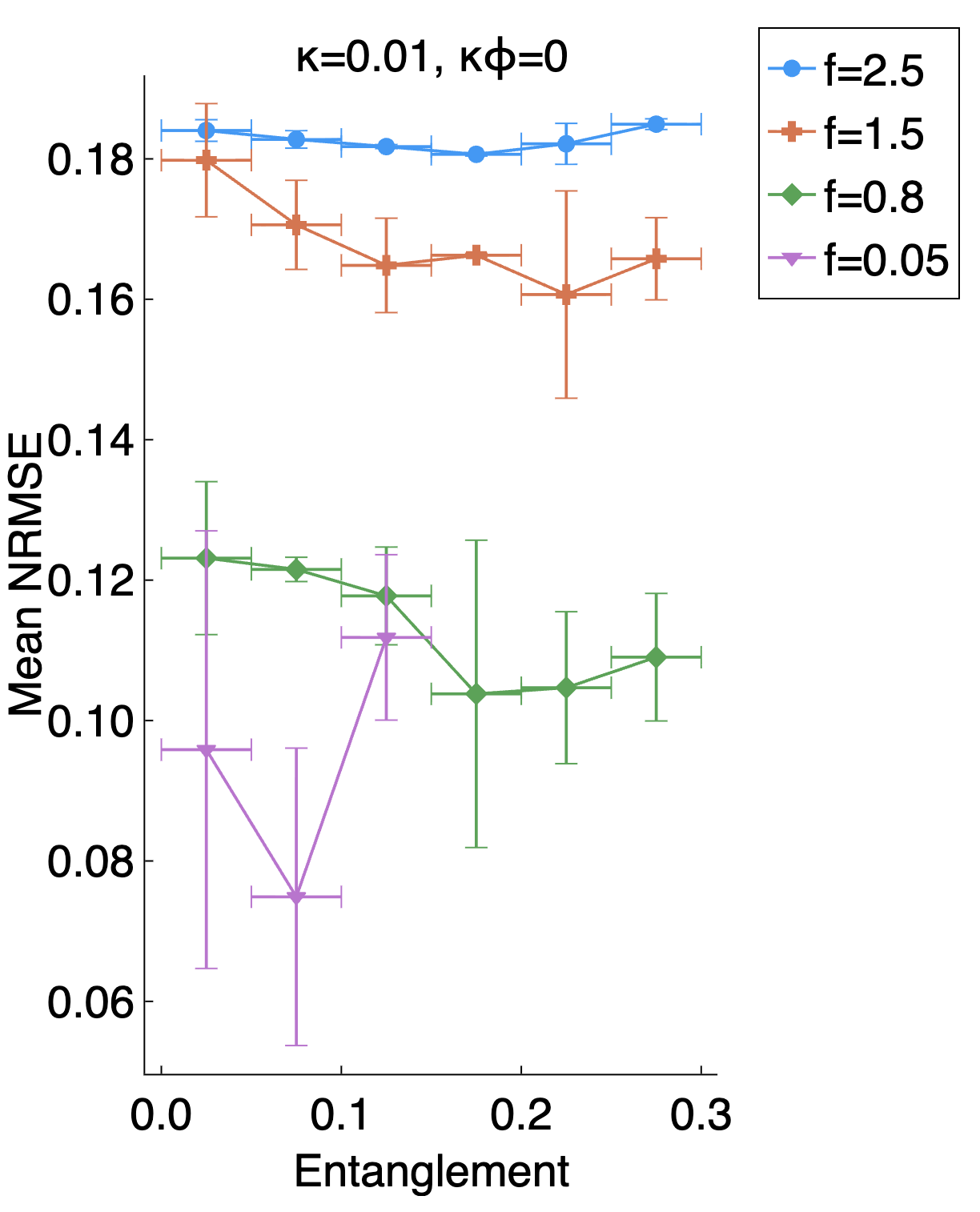} &
    \includegraphics[width=0.41\textwidth]{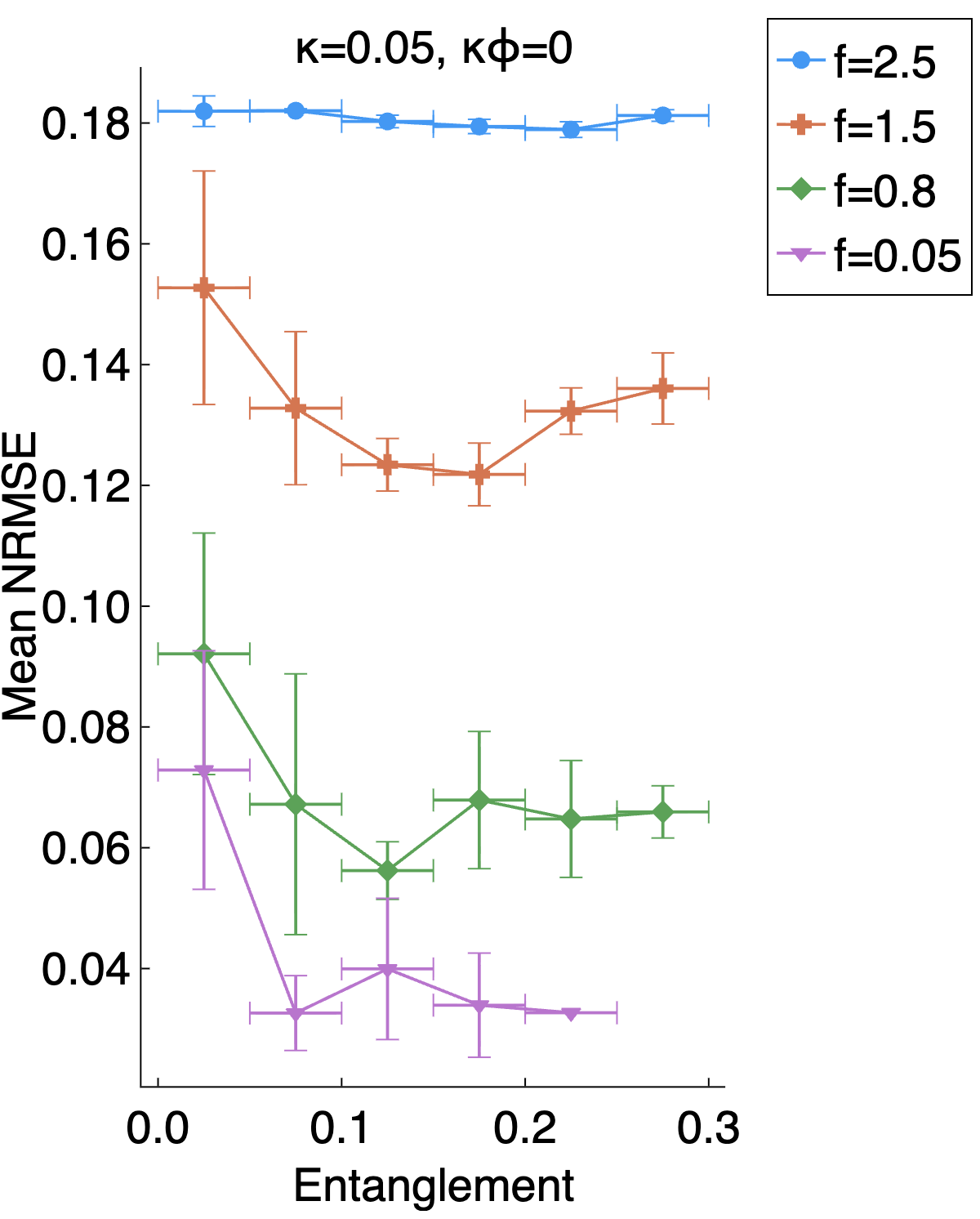} \\
    \includegraphics[width=0.35\textwidth,trim={0 0 {.27\textwidth} 0},clip]{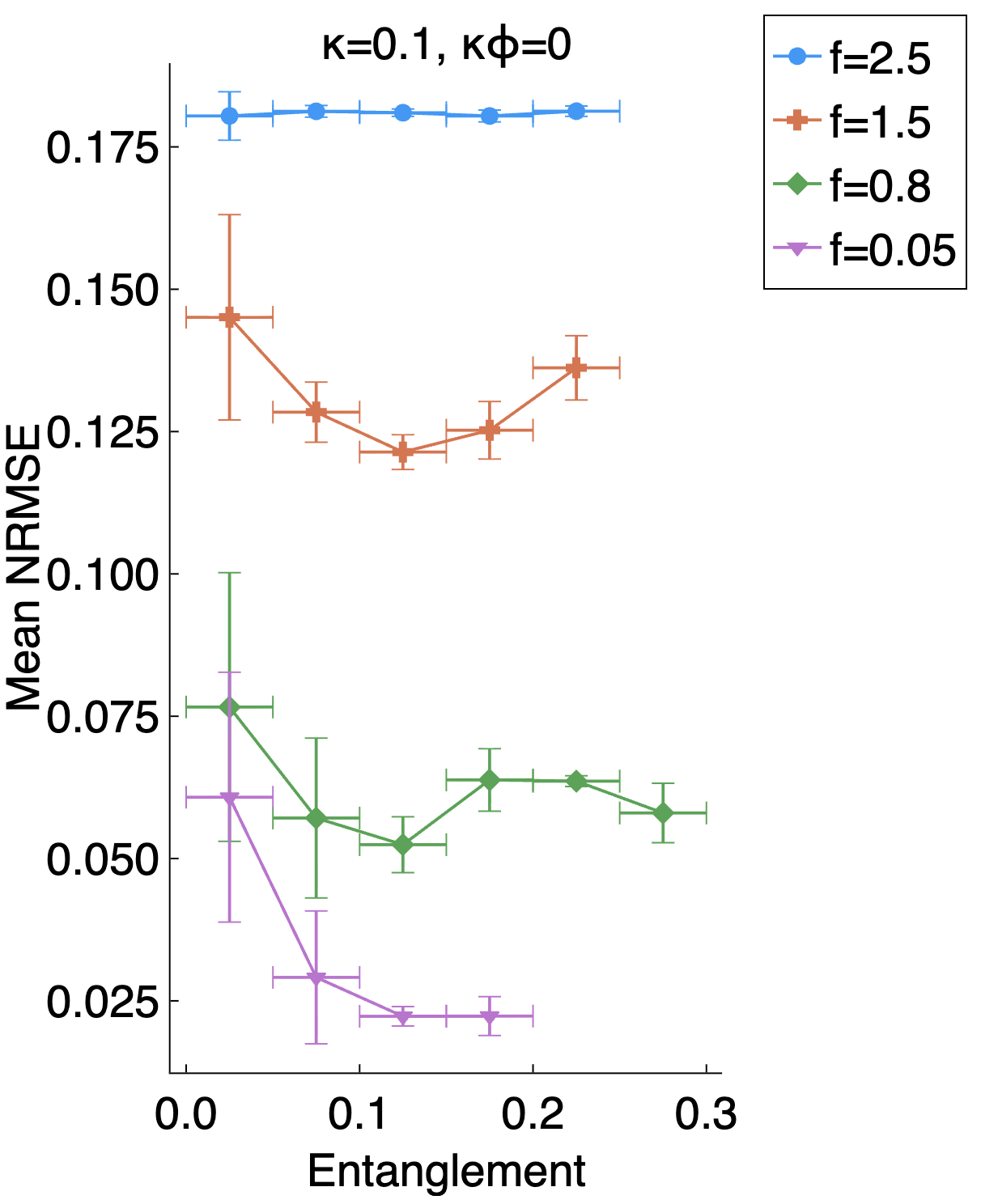} &
    \includegraphics[width=0.45\textwidth]{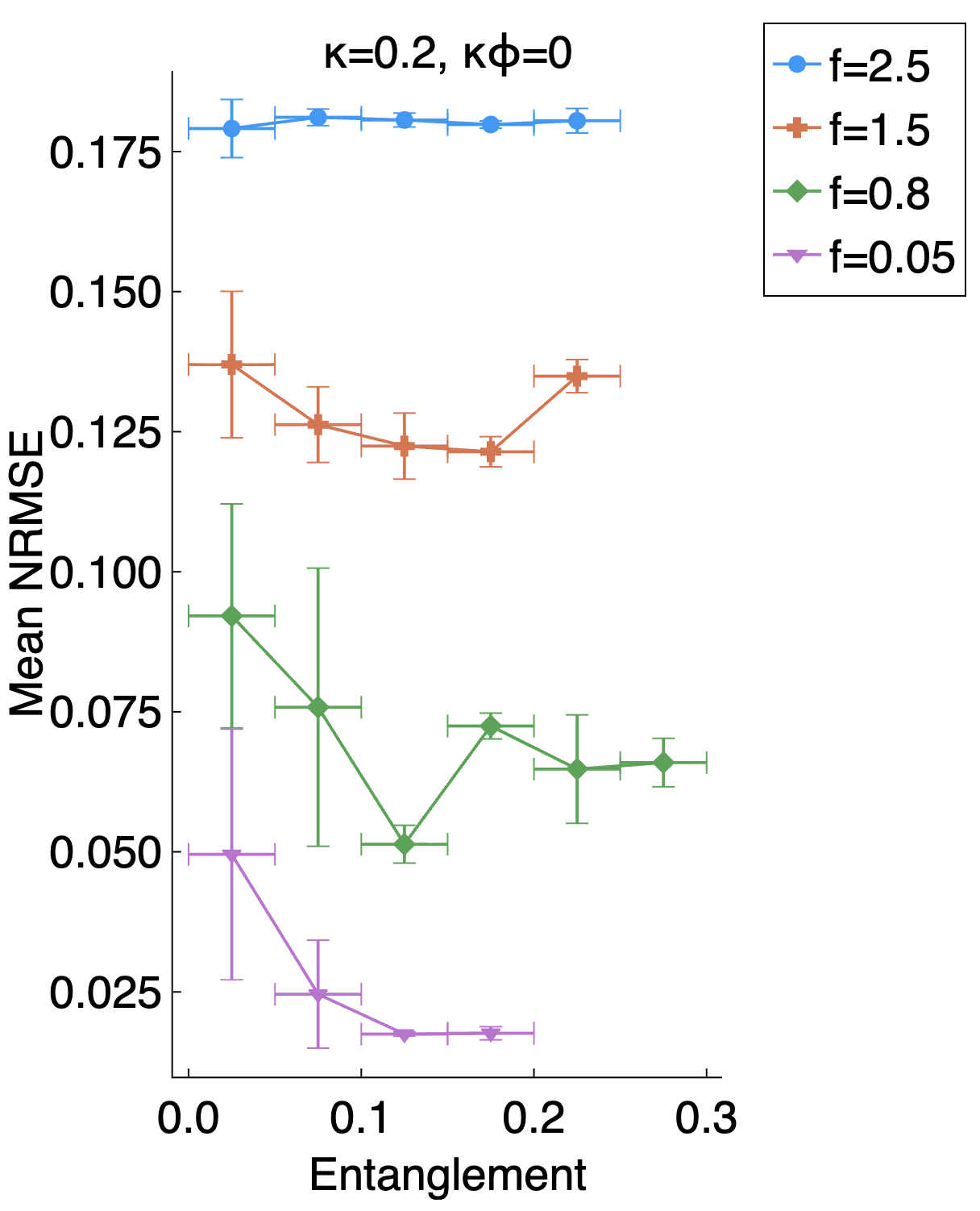}\\
  \end{tabular}
  \caption{Same as Fig.~\ref{fig:ent} but for different dissipations, $\kappa$.}
  \label{fig:pk}
\end{figure*}
Fig.~\ref{fig:pk} shows the full binned relation between NRMSE and entanglement across multiple dissipation rates. The trend confirms that for sufficient dissipation, moderate entanglement is associated with better prediction performance, up to a saturation point. This supports our claim that optimal entanglement lies in an intermediate but non-zero regime.

\begin{figure*}[htbp]
  \centering
  \begin{tabular}{ccc}
    \includegraphics[width=0.3115\textwidth,trim={0 0 {.27\textwidth} 0},clip]{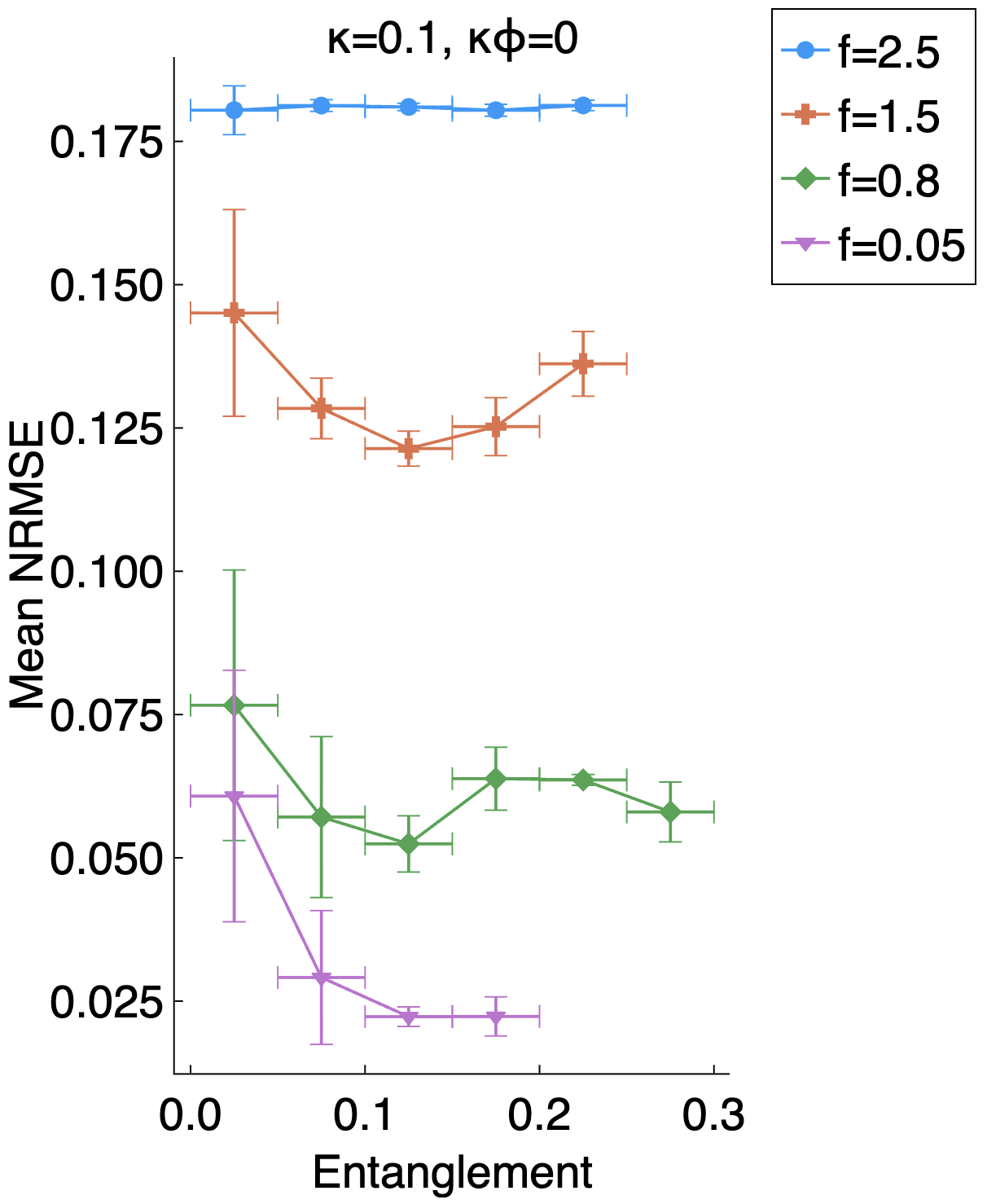} &
    \includegraphics[width=0.4\textwidth]{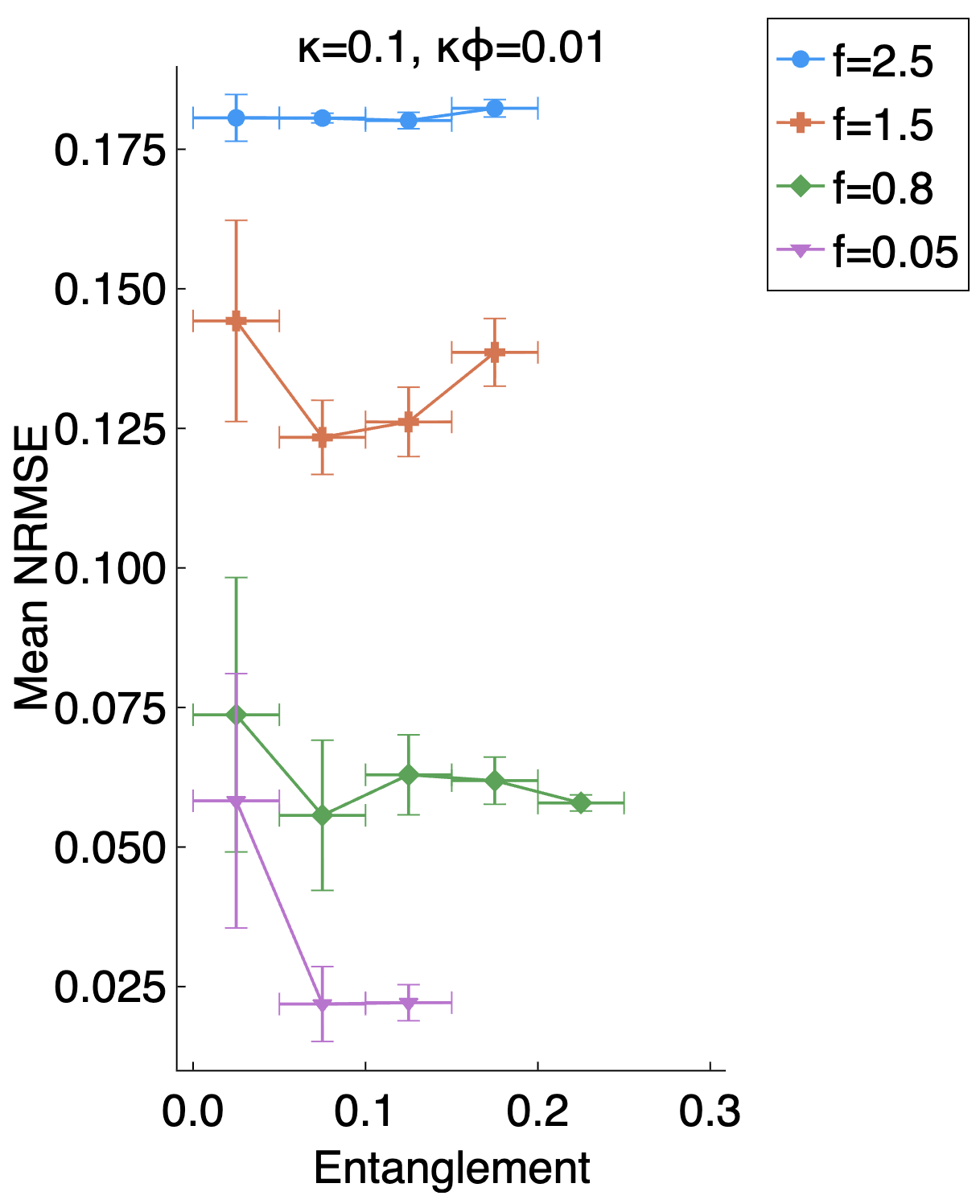} \\
    \includegraphics[width=0.3115\textwidth,trim={0 0 {.27\textwidth} 0},clip]{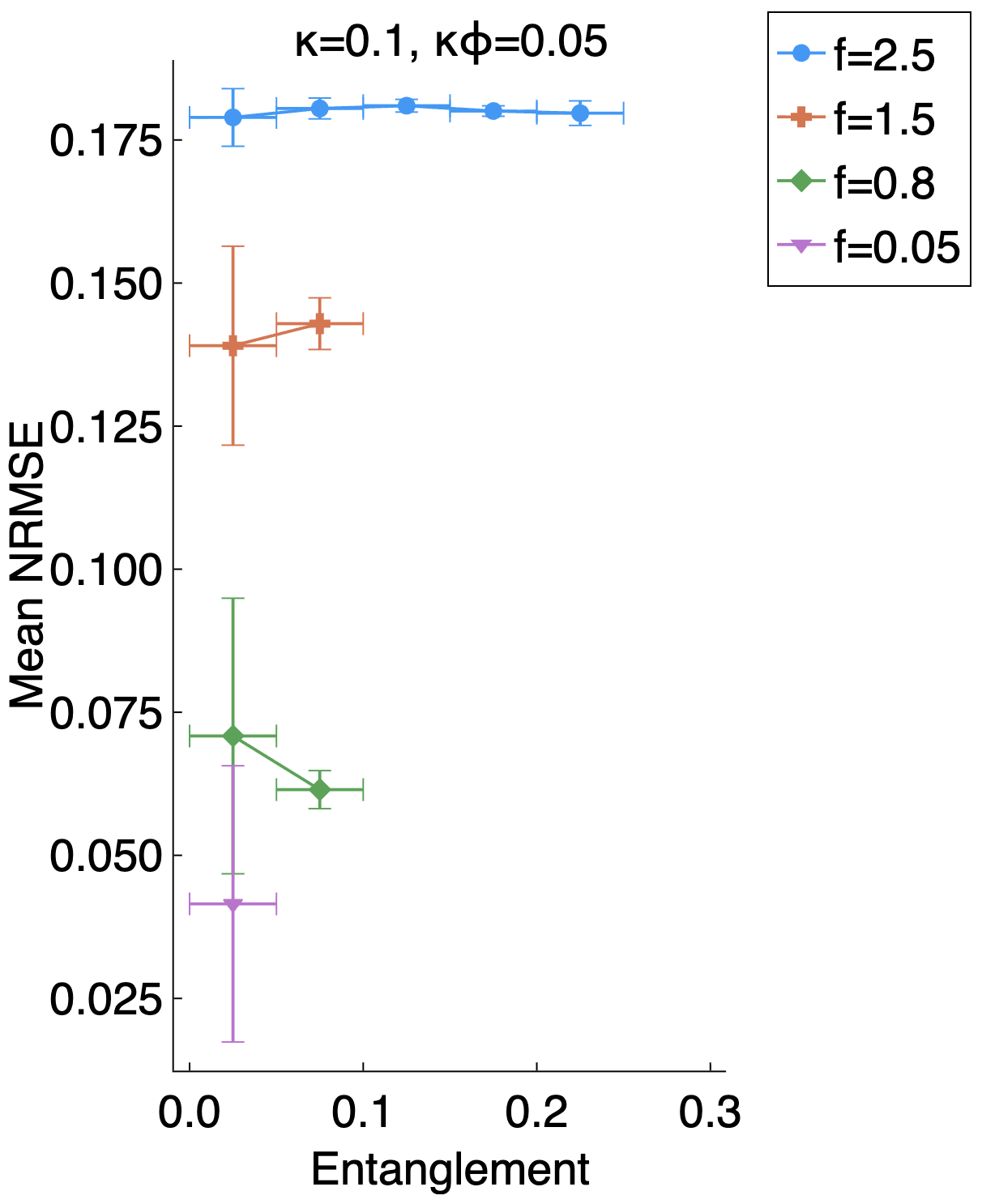} &
    \includegraphics[width=0.4\textwidth]{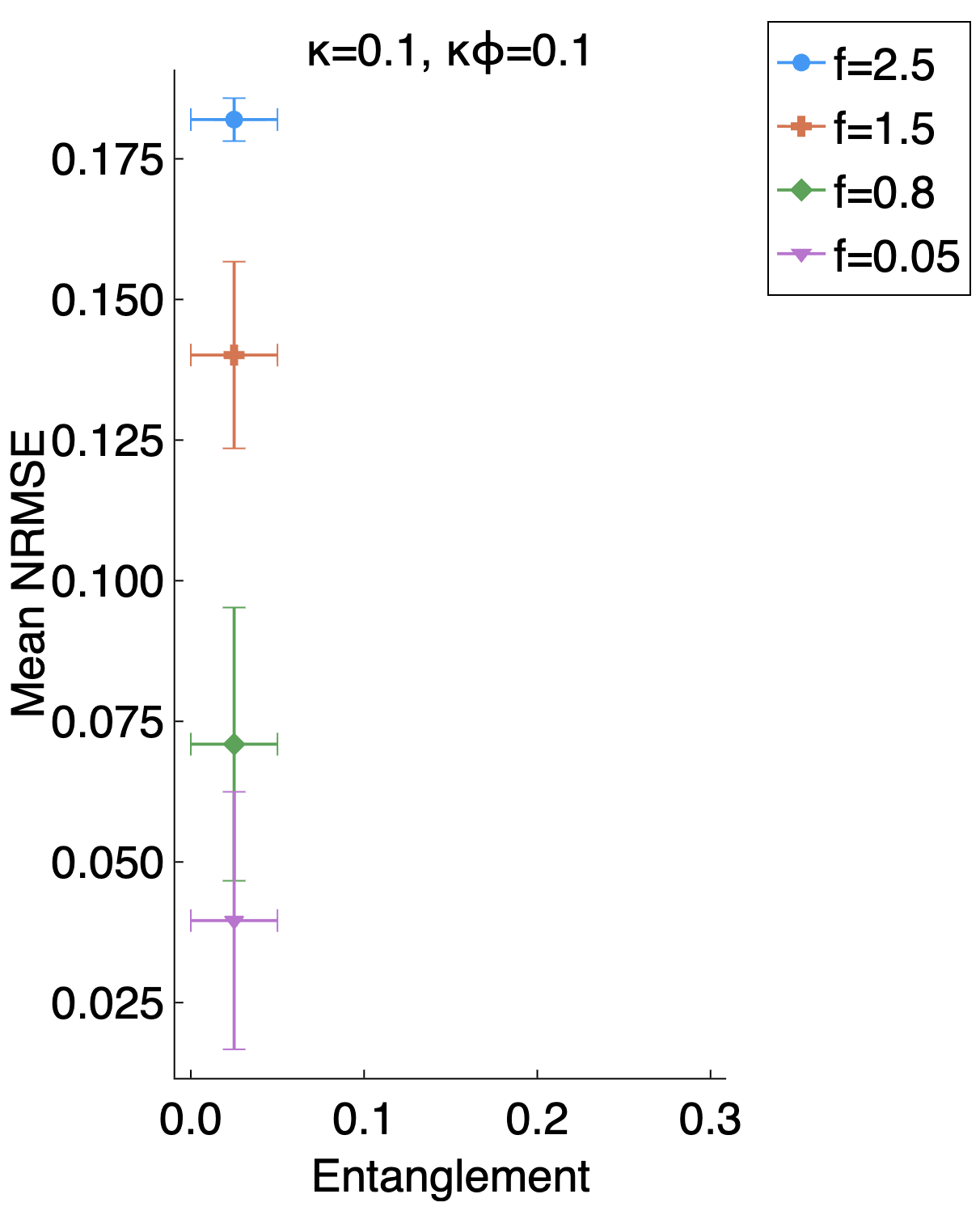}\\
  \end{tabular}
  \caption{ Same as Fig.~\ref{fig:ent} but for different dephasings, $\kappa_\phi$.}
    \label{fig:pd}
\end{figure*}

Fig.~\ref{fig:pd} provides a similar analysis for various dephasing rates. The best results in most cases are obtained when $\kappa_\phi = 0$. While small nonzero values can provide marginal benefit at low frequencies, performance generally declines as dephasing increases.

\section{Cutoff Dimension}
\label{app:cut}
\begin{figure}[htbp]
  \centering
  \includegraphics[width=0.4\textwidth]{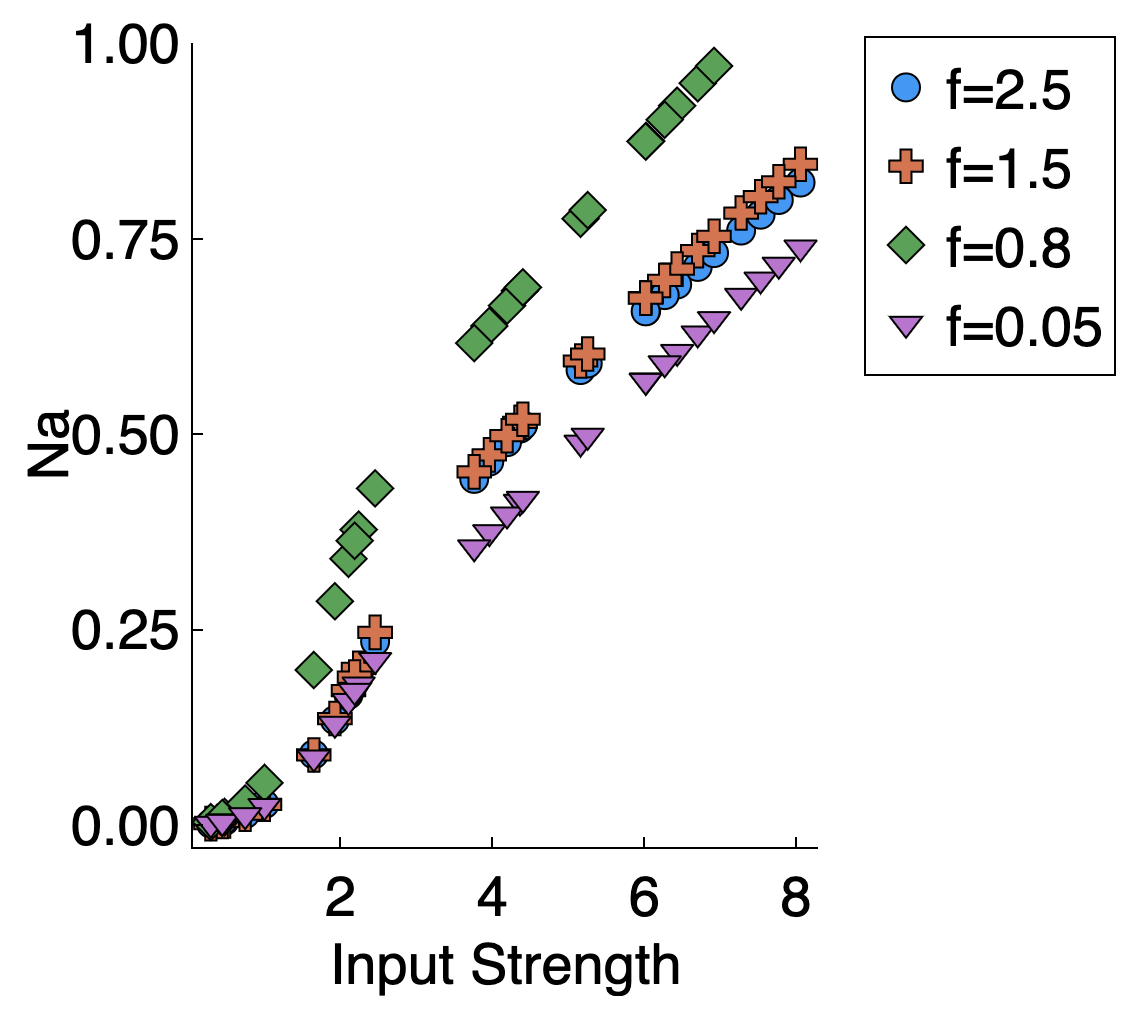}
  \caption{Phonon number Vs different input strengths for different frequencies. The phonon number increases as the input strength increases but remains below 1.}
  \label{fig:cutoff}
\end{figure}

We impose an average excitation number cutoff  $N_{\text{cut}}$  on each mode, truncating the Fock basis to states  $\ket{n_a, n_b}$  with  $n_a$, $n_b$ $\leq N_{\text{cut}}$ . This cutoff is necessary for numerical feasibility but introduces an artificial boundary. To ensure the results are physically meaningful and independent of the cutoff, the average excitation number populations must remain well within this truncated space specially when we increase the input strength. To that end, we constrain the mean excitation  numbers $\langle \hat{N}_a \rangle$ and $\langle \hat{N}_b \rangle$ to remain below one (Fig.~\ref{fig:cutoff}).
\section{NARMA Task}\label{app:narma}
\subsection{Entanglement-Performance Relationship}
To assess the reservoir's capability on a complex nonlinear task, we employed the NARMA-20 kernel~\cite{Rodannarma}. Crucially, rather than using the standard white noise input often utilized for memory capacity estimation, we applied the NARMA-20 recurrence relation to the same deterministic multi-frequency signal ($s_k$) used in our main analysis. This approach allows us to stress-test the reservoir against a highly nonlinear dynamical rule. The series is defined by:
\begin{equation}
\begin{split}
    y_{t+1} = \tanh \Big( & 0.3 y_t + 0.05 y_t \sum_{j=0}^{19} y_{t-j} \\
    & + 1.5 u_{t-20} u_{t-1} + 0.1 \Big) + 0.2
\end{split}
\label{eq:narma20}
\end{equation}

where $y_t$ represents the generated time-series using the NARMA-20 kernel, and $u_t$ corresponds to the input signal. Crucially, in this work, $u_t$ is not a random variable but is set to equal the multi-frequency signal ($s_k$) used in the main text. We implemented an offline NARMA kernel to the multi-frequency input signal and then did a prediction task on the resulting time-series, $y_t$, using half of it for training and half of it for testing just like what we did for the main task.

\begin{figure}[htbp]
  \centering
  \begin{tabular}{ccc}
    \includegraphics[width=0.21\textwidth,trim={0 0 {.27\textwidth} 0},clip]{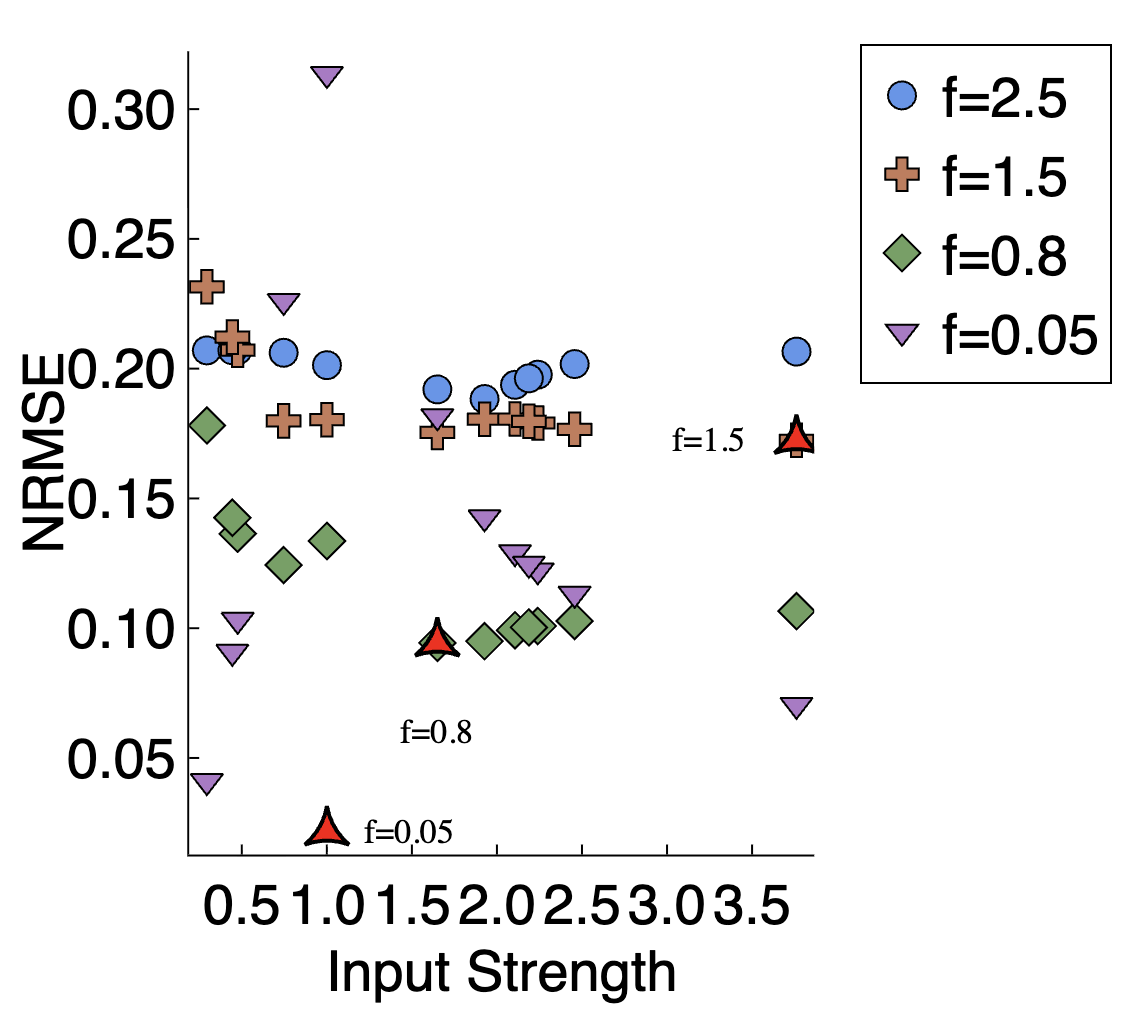} &
    \includegraphics[width=0.26\textwidth]{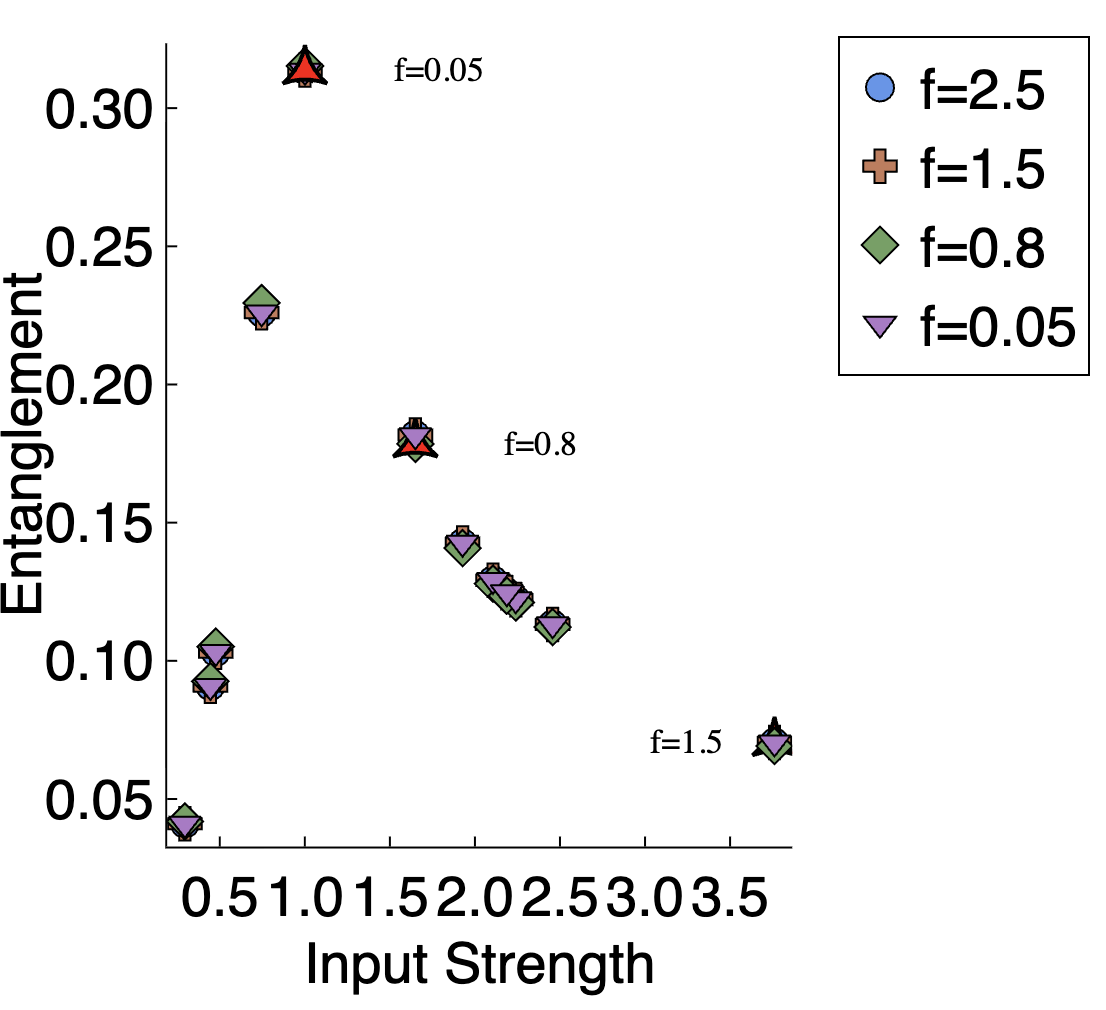} \\
  \end{tabular}
  \caption{Same as Fig. \ref{fig:inpst}  for the NARMA task. }
  \label{fig:narm1}
\end{figure}

\begin{figure}[htbp]
  \centering
  \begin{tabular}{ccc}
    \includegraphics[width=0.21\textwidth,trim={0 0 {.27\textwidth} 0},clip]{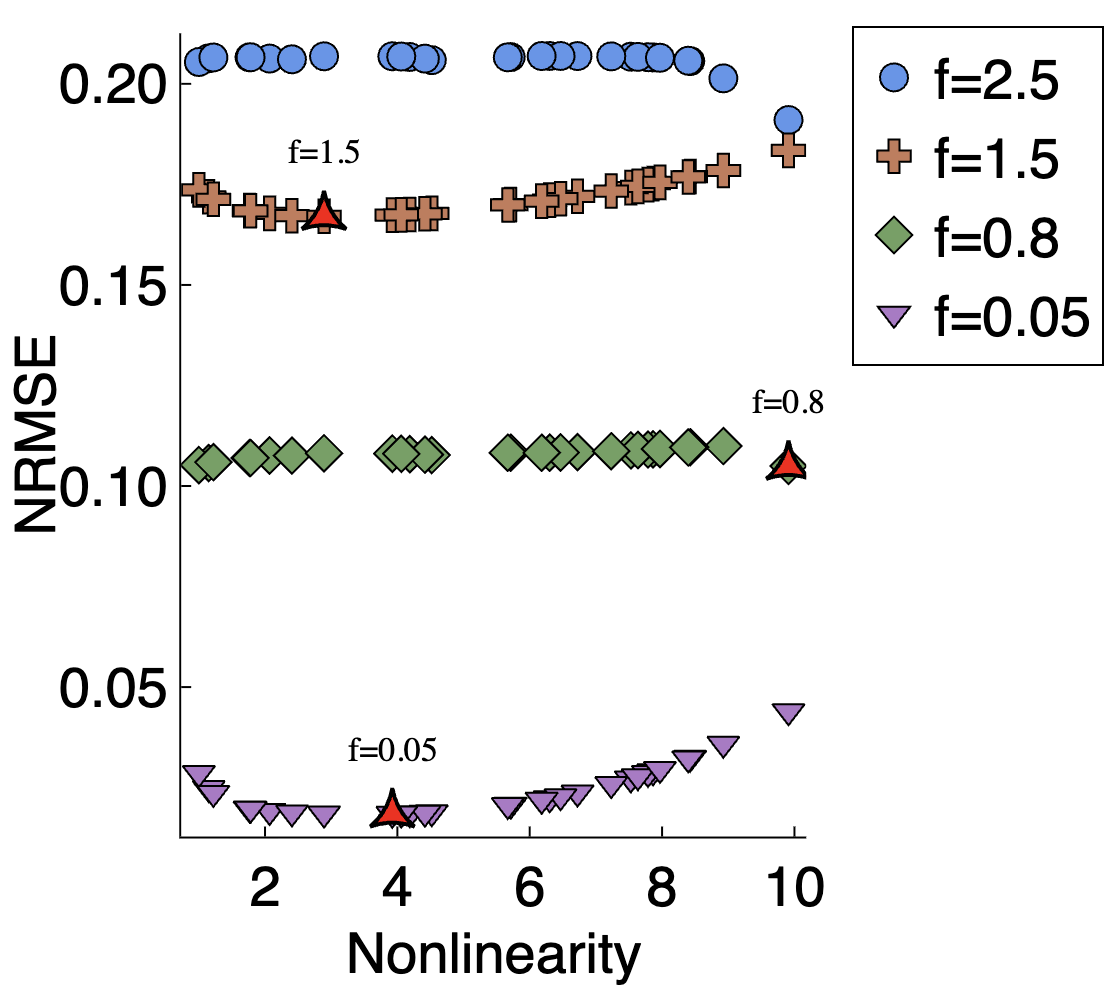} &
    \includegraphics[width=0.27\textwidth]{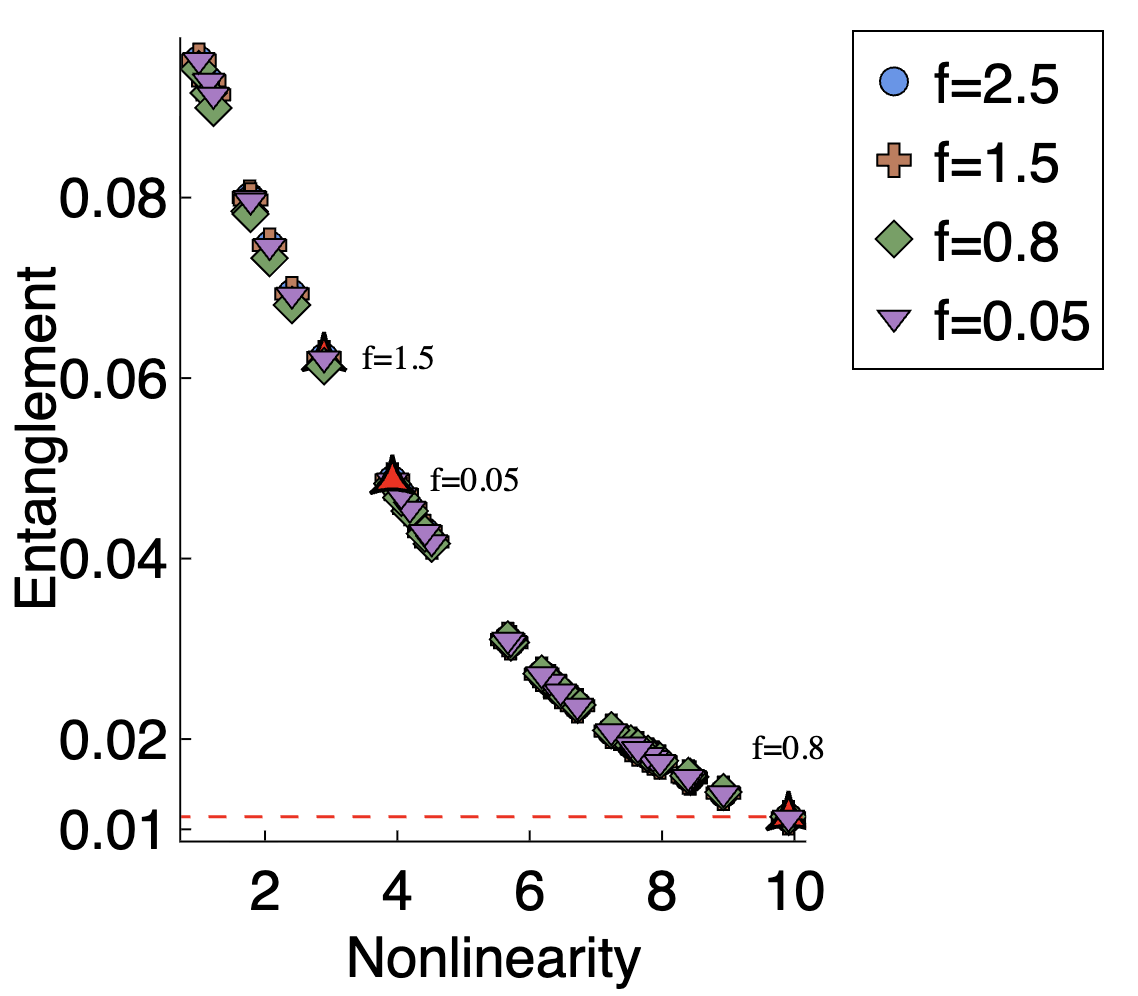} \\
  \end{tabular}
  \caption{Same as Fig. \ref{fig:knl} but for the NARMA task. }
  \label{fig:narm2}
\end{figure}

\begin{figure}[htbp]
  \centering
  \begin{tabular}{ccc}
    \includegraphics[width=0.2\textwidth,trim={0 0 {.27\textwidth} 0},clip]{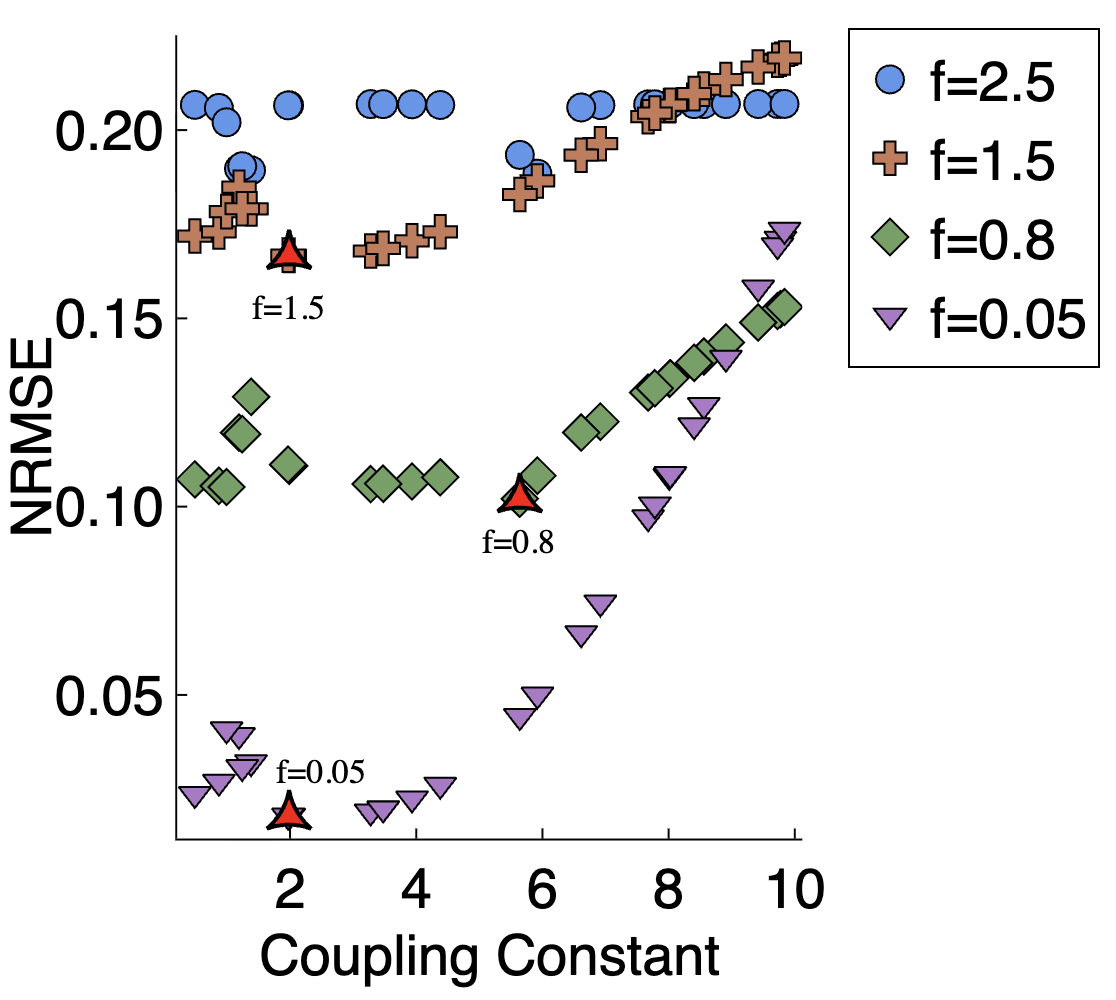} &
    \includegraphics[width=0.26\textwidth]{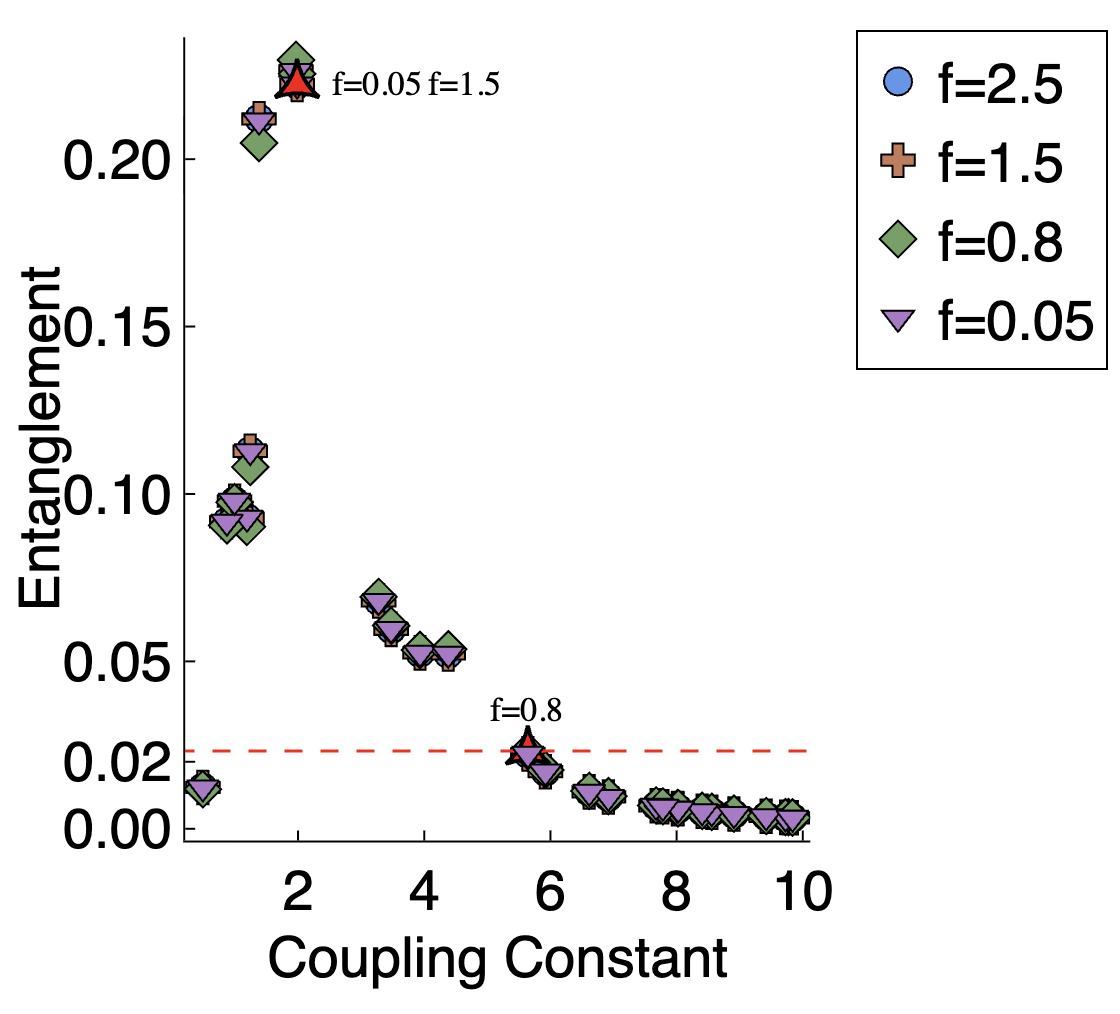} \\
  \end{tabular}
  \caption{Same as Fig. \ref{fig:cc} but for the NARMA task. }
  \label{fig:narm3}
\end{figure}

We repeated the parameter sweeps performed for the linear task with the same conditions (such as the limit on the mean excitation number) to verify the robustness of our findings. Figures \ref{fig:narm1}, \ref{fig:narm2}, and \ref{fig:narm3} display the NRMSE and entanglement behavior under varying input strength, nonlinearity, and coupling, respectively. 

\begin{figure}[htbp]
    \centering
        \includegraphics[width=0.29\textwidth]{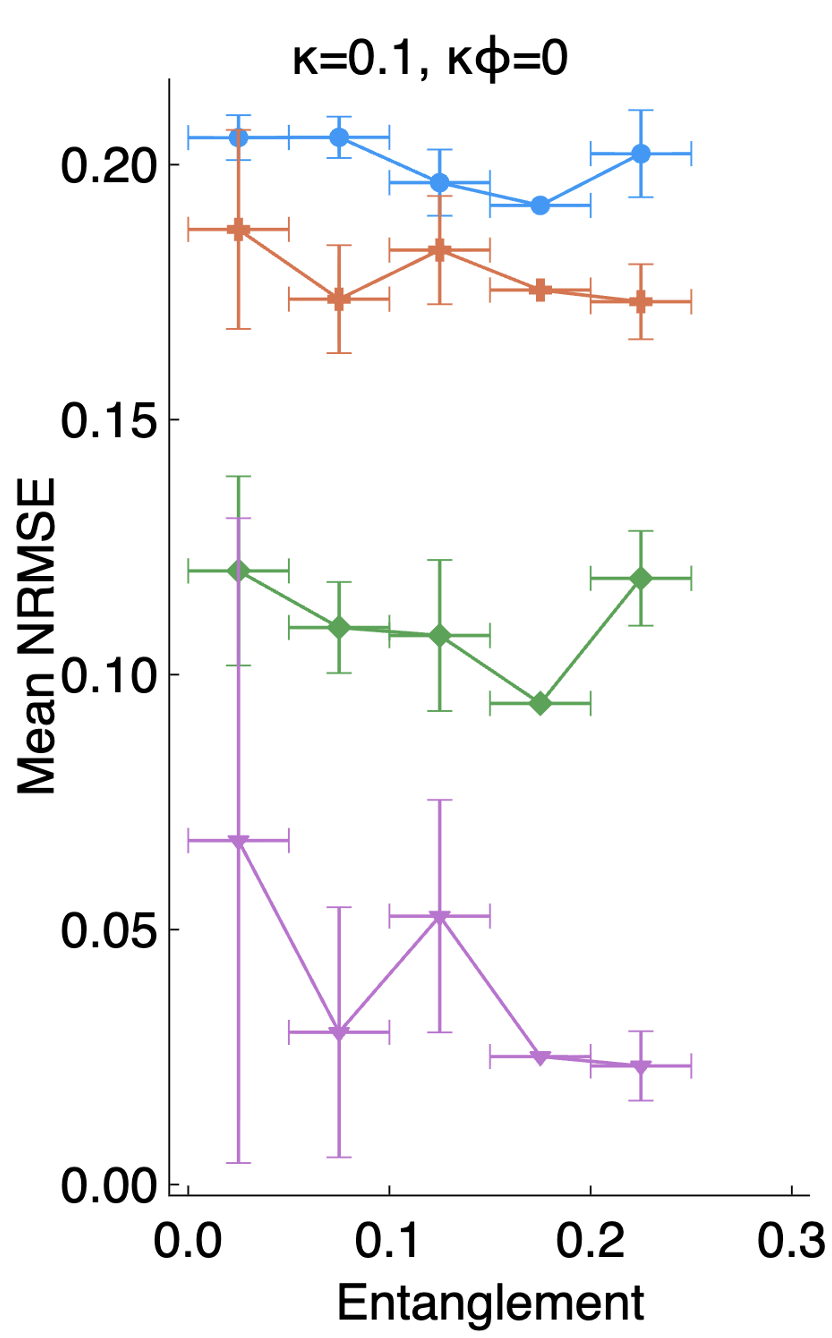}
    \caption{Same as Fig. \ref{fig:ent} but for the NARMA task.}
    \label{fig:narm4}
\end{figure}

Notably, we consolidated the results from the Figures \ref{fig:narm1}, \ref{fig:narm2}, and \ref{fig:narm3}, and performed a binned analysis to produce Fig.~\ref{fig:narm4} which demonstrates that the relationship between entanglement and predictive performance remains consistent with the linear task results: moderate but non-zero entanglement corresponds to minimum errors. This confirms that the observed entanglement-performance relationship is not an artifact of the specific linear task used in the main text but persists in a more complex and nonlinear forecasting benchmark.
\subsection{Dissipation \& Dephasing}
Here we do the same procedure to produce Figures \ref{fig:diss} and \ref{fig:deph} but for the NARMA-20 time-series prediction task described above.

\begin{figure}[htbp]
    \centering
        \includegraphics[width=0.8\linewidth]{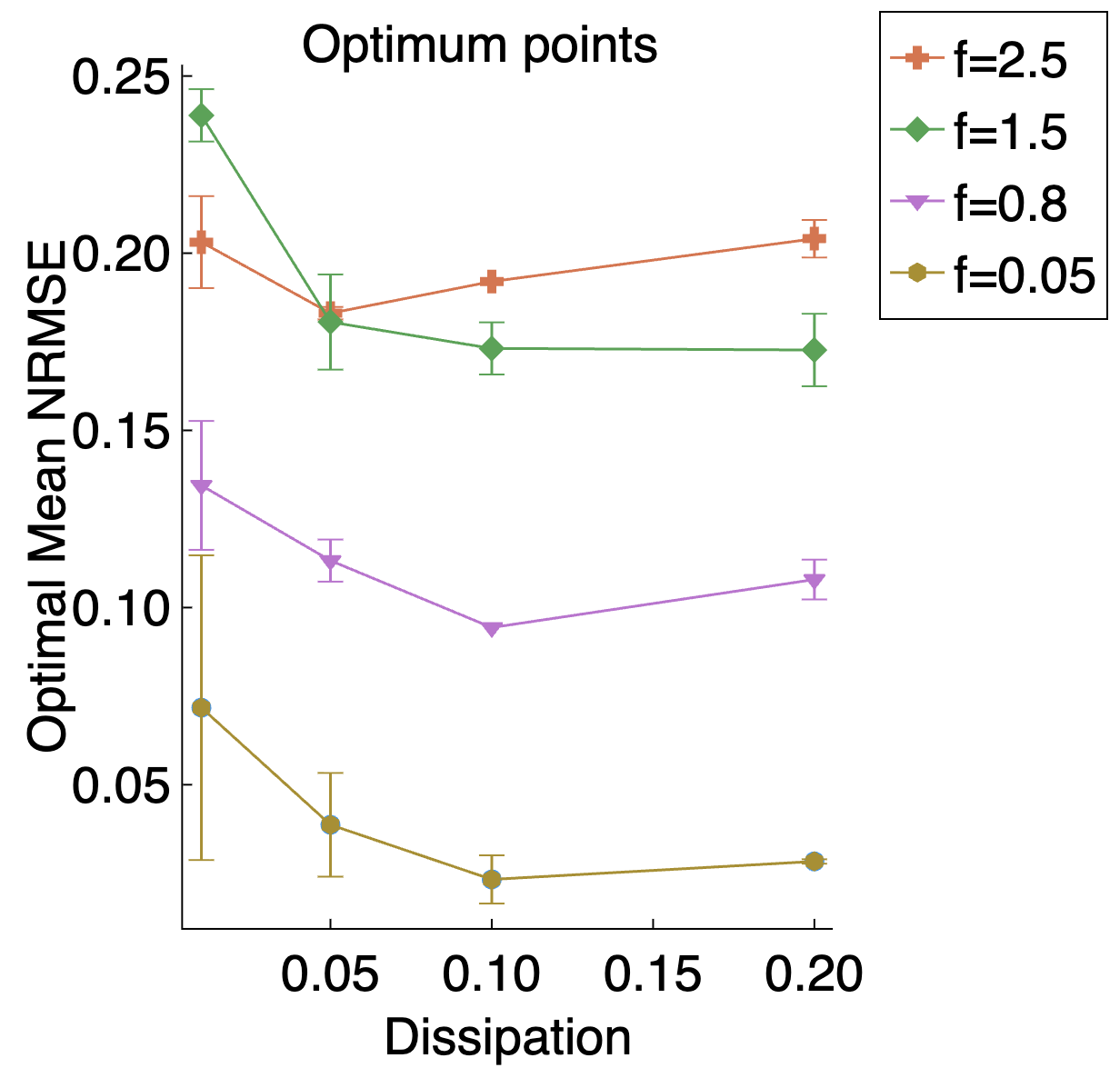} 
        \caption{Same as Fig. \ref{fig:diss} but for NARMA task}
    \label{fig:narmds}
\end{figure}

As shown in Fig.~\ref{fig:narmds}, increasing the dissipation rate yields a clear improvement in predictive performance across all valid frequency scales ($f \le 1.5$), with an optimal operating point appearing near $\kappa \approx 0.1$ 
(except for $f=2.5$). This behavior mirrors the findings in the linear task, confirming that a sufficient level of dissipation is necessary.

\begin{figure}[htbp]
    \centering
        \includegraphics[width=0.8\linewidth]{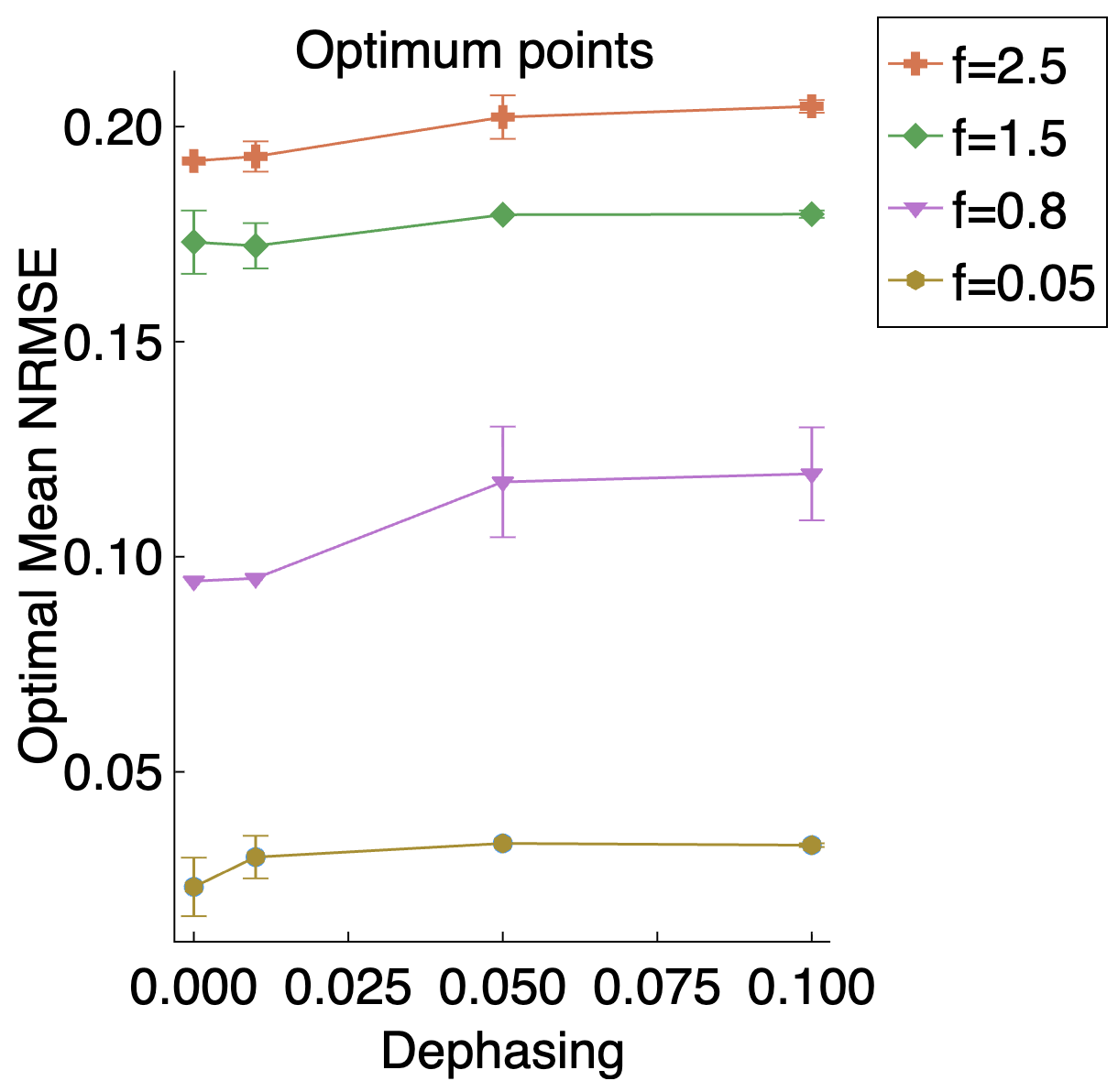} 
        \caption{Same as Fig. \ref{fig:deph} but for NARMA task.}
        \label{fig:narmdf}
\end{figure}

In contrast, Fig.~\ref{fig:narmdf} demonstrates that dephasing is generally detrimental to performance. The minimum error is consistently achieved at $\kappa_{\phi} = 0$ for all frequencies. Unlike the linear task, where a small amount of dephasing provided a marginal benefit at the lowest frequency ($f=0.05$), the NARMA task shows an immediate degradation in error as dephasing is introduced for all the frequencies.
    
    


\clearpage

\bibliography{main}

\end{document}